\documentclass[superscriptaddress,twocolumn,showpacs,preprintnumbers,prb]{revtex4}
\usepackage{graphicx}
\usepackage{times}

\usepackage{color}

\begin{document}

\def\salto{\vskip 1cm}
\def\lgr{\langle\langle}
\def\rgr{\rangle\rangle}

\title{Transport in Carbon Nanotubes: 2LSU(2) regime reveals subtle competition between Kondo and Intermediate Valence states}

%
\author{C.~A. B\"usser}
\affiliation{Department of Physics, Oakland University, Rochester, MI 48309, USA}
\affiliation{Dept. of Physics and Astronomy, University of Wyoming, Laramie, WY 82071, USA}
\author{E. Vernek}
\affiliation{Instituto de F\'isica - Universidade Federal de Uberl\^andia - Uberl\^andia, 
MG 38400-902 - Brazil}
\author{P. Orellana}
\affiliation{Departamento de F\'{\i}sica, Universidad Cat\'olica del Norte, Casilla 1280, Antofagasta, Chile}
\author{G.~A. Lara}
\affiliation{Departamento de F\'{\i}sica, Universidad de Antofagasta, Casilla 170, Antofagasta, Chile}
\author{E.~H. Kim}
\affiliation{Department of Physics, University of Windsor, Windsor, ON, N9B 3P4, Canada}
\author{A.~E. Feiguin}
\affiliation{Dept. of Physics and Astronomy, University of Wyoming, Laramie, WY 82071, USA}
\author{E.~V. Anda}
\affiliation{Departamento de F\'{\i}sica, Pontif\'{\i}cia Universidade Cat\'olica do Rio de Janeiro, 22453-900, Brazil.}
\author{G.~B. Martins}
\email[Corresponding author: ]{martins@oakland.edu}
\affiliation{Department of Physics, Oakland University, Rochester, MI 48309, USA}
%

\begin{abstract}
{In this work, we use three different numerical techniques to study the charge transport properties of
a system in the two-level SU(2) (2LSU2) regime, obtained from
an SU(4) model Hamiltonian by introducing orbital mixing of the degenerate orbitals via coupling to the leads.
SU(4) Kondo physics has been experimentally observed, and studied in detail, in Carbon Nanotube Quantum Dots. 
Adopting a two molecular orbital basis, the Hamiltonian is recast into
a form where one of the molecular orbitals decouples from the charge reservoir, although still interacting
capacitively with the other molecular orbital. 
This basis transformation explains in a clear way how the charge transport in this system turns from double- to single-channel 
when it transitions from the SU(4) to the 2LSU2 regime. 
The charge occupancy of these molecular orbitals displays gate-potential-dependent occupancy
oscillations that arise from a competition between the Kondo and Intermediate Valence states.
The determination of whether
the Kondo or the Intermediate Valence state is more favorable, for a specific value of gate potential, is assessed
by the definition of an energy scale $T_0$, which is calculated through DMRG.  
We speculate that the calculation of $T_0$ may provide experimentalists with a useful tool to
analyze correlated charge transport in many other systems. For that, a current work is underway 
to improve the numerical accuracy of its DRMG calculation and explore different definitions.
}
\end{abstract}

\maketitle

\section{Introduction} 
Interest in charge transport properties of carbon nanotubes (CNT)\cite{saito} has grown consistently 
in the last few years, in part because of their possible applications as 
single electron transistors (SETs),\cite{dekker1} as well as their possible use in molecular computing.\cite{lieber}
A CNT SET can be manufactured by coupling a CNT quantum dot (QD) to metal leads\cite{dekker1}. 
In this case, below a certain characteristic temperature $T_K$, the Kondo effect\cite{hewsonbook} 
was first observed in CNT QDs in 2000.\cite{kondocnt} 
Further investigation showed that the main difference between the Kondo effect in 
CNT QDs, and the same effect in other nano-structures, comes 
from the degeneracy of the orbital (chiral) states of the CNT QD.\cite{kondosu4} Effectively, while in the one hand the Kondo effect in, e.g., 
a lateral semiconducting QD is associated to the screening of a localized magnetic moment (in general, with SU(2) symmetry) 
by the spin density of the conduction electrons, in the other hand, in a CNT QD the two degenerate orbital 
states with different chirality (usually referred to as a pseudospin), together with the intrinsic spin degree 
of freedom, give rise to the so-called SU(4) Kondo effect. 
This exotic state has been recently studied both experimentally \cite{kondosu4} and theoretically. \cite{SU4-theo,logan,Aguado,martins1} 
Some of the many particular characteristics of the SU(4) Kondo effect are (i) a larger Kondo temperature than in the SU(2) case, (ii) Kondo effects at 
two different fillings, namely, quarter-filling and half-filling, with diverse properties,\cite{martins1} (iii) a rich behavior 
under magnetic field,\cite{eugene} (iv) perfect entanglement of spin and orbital degrees of freedom,\cite{kondosu4} among others.

Another interesting regime in the SU(4) Kondo Hamiltonian can be obtained by the 
gradual `mixing' of orbital states at the tunneling barriers, through tunneling that 
effects a flip in the pseudospin of the electron [henceforth, pseudospin conserving 
tunneling occurs through a matrix element $t^{\prime}$, and non-conserving through $t^{\prime \prime}$, see Fig.~1(a)]. 
When this mixing is total (i.e., $t^{\prime \prime}/t^{\prime} = 1$), one 
reaches the so-called two-level SU(2) (2LSU2) state.\cite{Aguado,martins1} 
One constraint associated to the use of a CNT QD to observe SU(4) Kondo physics 
is the lack of control over the parameters of the system 
(intra- and inter-orbital Coulomb repulsions $U$ and $U^{\prime}$, respectively, and couplings to leads), making 
it impossible to freely navigate from the SU(4) to the 2LSU2 state. 
In contrast, recent advances in the lithography of semiconducting lateral 
QDs have allowed greater control over capacitively coupled double quantum dot (DQD) systems
connected to independent leads.\cite{2lsu2-exp} In reality, it was in a DQD system that the 
first claim of observation of the SU(4) Kondo effect was made.\cite{first-exp} 
As long as progress in this field continues to be made, it is not unlikely that the
2LSU2 regime will be accessible to experimental probing in the near future. Indeed, in the same way that the SU(4) regime has been
shown to be quite robust regarding a difference between $U$ and $U^{\prime}$,  and 
some mixing of the orbital states (although {\it strict} SU(4) symmetry only occurs 
at one point in parameter space), \cite{logan} most of the relevant phenomena described here for the 2LSU2 state 
are shown to survive a deviation of $t^{\prime \prime}/t^{\prime}$ from $1$. It is for that reason 
that we elect to use the terminology SU(4) and 2LSU2 {\it regimes}, since as far as experiments go, 
their characteristic properties do not seem to be restricted to the corresponding point in parameter 
space (we choose to refer to these specific points as SU(4) and 2LSU2 {\it states}). 

In this paper, using three different numerical techniques, namely the 
Numerical Renormalization Group (NRG),\cite{theo} the Density 
Matrix Renormalization Group (DMRG),\cite{white} and the Logarithmically Discretized 
Embedded Cluster Approximation (LDECA),\cite{ldeca} we analyze in detail the opposite end of 
the SU(4) regime, the above mentioned 2LSU2 regime. \cite{Aguado,martins1}
Using the notation defined above, the difference between the SU(4) and 2LSU2 regimes 
resides in the value of the ratio $\nu=t^{\prime \prime}/t^{\prime}$: 
$\nu=0$ corresponds to the SU(4) regime (no orbital mixing) and $\nu=1$ to the 2LSU2 regime (total orbital mixing). 
The motivations for the study presented here are four-fold: First, in previous 
work by two of the current authors, a charge discontinuity and a transition from a $2G_0$ to 
$G_0$ maximum conductance was observed in the 2LSU2 regime ($G_0$ is the conductance quantum).\cite{martins1,martins2} 
Second, in these same works, a fine-structure of 
conductance peaks ($2G_0$ high and positioned at and around the particle-hole (p-h) symmetric point) was observed for 
parameter values in the 2LSU2 regime, superimposed over a flat background of height $G_0$. 
Third, a slave boson analysis of the local density 
of states (LDOS) showed a discontinuous change of the Kondo peak {\it at} the 2LSU2 regime. \cite{Aguado} 
All these three points mentioned above were never fully explained, some of them being thought to be 
related to convergence problems in the numerical methods used. \cite{martins3} Here, we will show that this is not 
the case, and that they are associated to a band-decoupling at the 2LSU2 state and to an interesting competition 
between two distinct many-body ground states. 
Finally, the above mentioned advances in manufacturing of capacitively coupled DQDs may open 
the doors for the experimental analysis of the regime theoretically discussed here. 
Note that to facilitate the connection and comparison with previous work by some of the 
authors, we will keep referring to the system under study as a CNT QD, although the 
regime of parameters to be analyzed ($\nu \lesssim 1$) may not be accessible to CNT QDs. 

Another interesting aspect of the results presented here is that the points 
mentioned above (charge discontinuity, change from double- to single-channel charge transport, fine structure of the conductance, 
and discontinuity in the LDOS Kondo peak) can be more easily understood after a 
change of basis from `atomic' to `molecular' orbitals (AOs to MOs) is applied. Indeed, QDs have 
been dubbed `artificial atoms', and systems of two or more coupled QDs have been dubbed `artificial molecules'. 
When the phenomena being analyzed depend more on the properties of quantum states involving more than one 
QD (a `molecular orbital'), a change of basis from local states (individual QDs) to extended states, 
involving two or more QDs, should be done to help understand the phenomena 
being observed. \cite{molecular} In this work, the authors will present a 
situation where this change of basis is crucial for the understanding of the numerical results. 

This work is divided as follows: Section II briefly presents a description of the model Hamiltonian, and 
the regime of parameters to be analyzed. In addition, a brief description of the band 
decoupling occurring at the 2LSU2 state is described by introducing molecular orbitals 
(a detailed description is provided in Appendix A). Section III presents charge vs. gate voltage results obtained 
with DMRG and NRG, at zero temperature. Charge oscillations in the MOs, as a function of gate voltage, 
are in perfect agreement between the two techniques. The NRG results are extended to finite temperatures, 
determining the energy scale of the phenomena relevant for the $V_g$-dependent charge oscillations. In addition, to facilitate the 
discussion of the results, the idea of an effective gate potential $V_g^{\ast}$ is introduced. 
Section IV presents LDECA results for charge, conductance, 
and spin- and charge-fluctuations, {\it away} from the 2LSU2 point, in good qualitative agreement with the DMRG results. 
Section V presents a discussion of the origin of the $V_g$-dependent charge oscillations observed in the MOs, and, more importantly, an explanation 
of its dependence on the coupling of the CNT QD to the leads. These charge oscillations in the MOs 
reveal a subtle competition between two many-body states, namely, the Kondo state and the Intermediate Valence 
state. A simple way of quantifying this competition is obtained by using DMRG simulations to calculate 
the energy $T_0$ of the system for a wide window of gate potential $V_g$ (see Eq.~(12) for its definition and discussion thereafter).  
In addition, the DMRG calculation of $T_0$ for the single impurity Anderson Model is compared to the 
Kondo temperature expression derived by Haldane.\cite{haldane} A finite size scaling of the DMRG results shows that $T_0$ 
tends to Haldane's expression in a narrow window around the p-h symmetric point, leading the 
authors to speculate that the definition and numerical calculation of $T_0$ may be helpful to 
study correlated charge transport for a broad region of gate potential, including the Kondo and Intermediate Valence 
regimes. Section VI presents our conclusions. 
Appendix A presents details of the basis transformation that results in the band decoupling essential
to understand the properties of the 2LSU2 regime, and Appendix B shows results for the non-interacting model 
($U=U^{\prime}=0.0$).

\begin{figure}
\centerline{\includegraphics[width=3.3in]{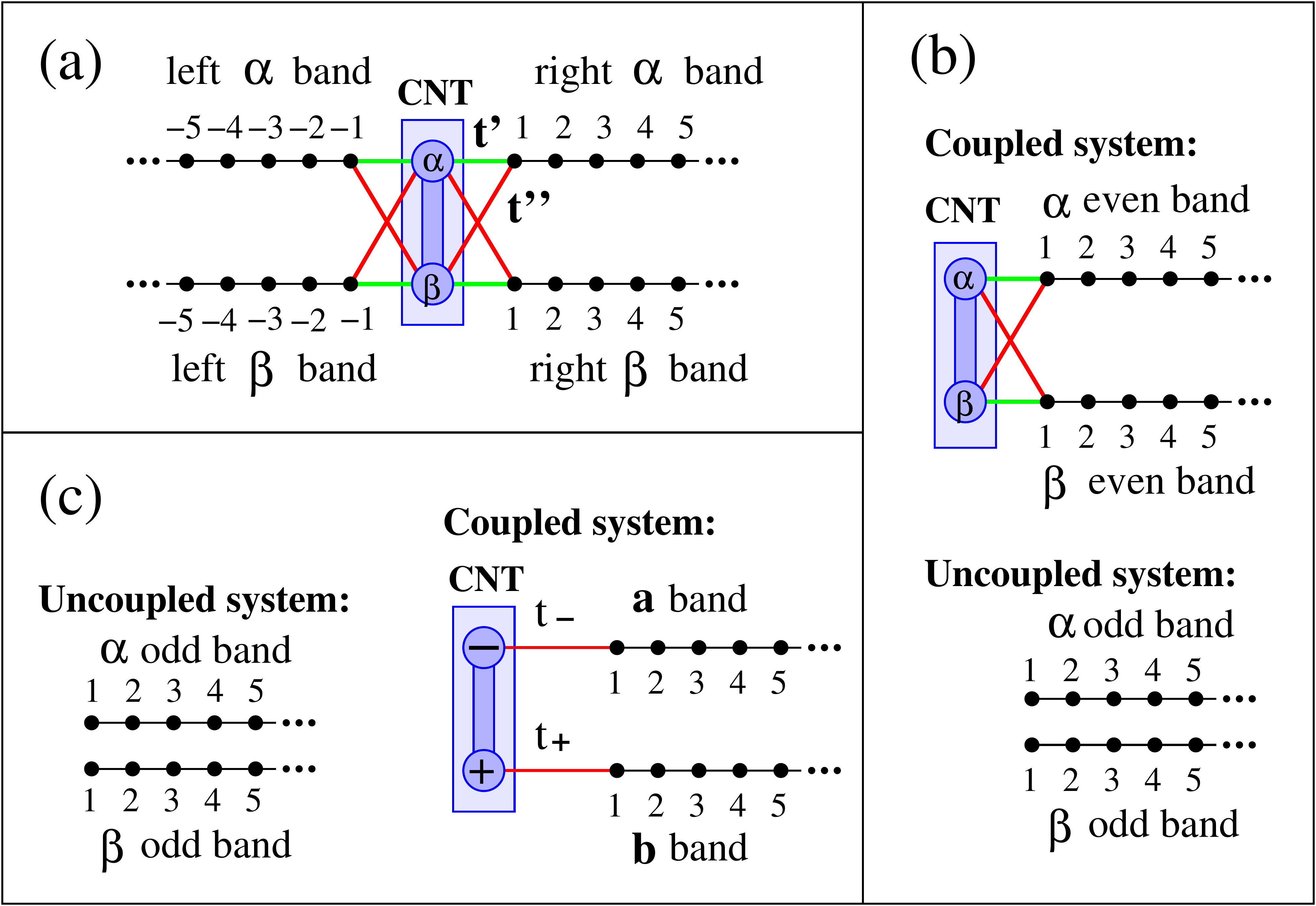}}
\caption{(a) Schematics of the {\it full} system studied
in this work, representing either a CNT (in which case $\alpha$ and $\beta$
stand for the orbital (chiral) levels of the CNT), or
a capacitively coupled DQD (in which case $\alpha$ and $\beta$ stand for two
QDs that are coupled to each other only through a long range Coulomb interaction).
In either case, the orbital levels of the QDs are each coupled
to its own independent charge reservoir, through hopping $t^{\prime}$
[(green) line]. Coupling to a reservoir with opposite quantum
number occurs through hopping $t^{\prime \prime}$ [(red) line].
The SU(4) state is achieved for $t^{\prime \prime}=0$,
and the 2LSU2 state is achieved for $t^{\prime \prime}=t^{\prime}$.
(b) Schematics of the usual left-right symmetry transformation that,
in this case, decouples two bands from the interacting levels.
(c) Final system, obtained after the second symmetry transformation (see text).
Now, the MOs $|+ \rangle$ and $|- \rangle$ are coupled to bands through
hoppings $t_+$ and $t_-$ that depend on the ratio $\nu = t^{\prime \prime}/t^{\prime}$.
For $\nu = 1$ (2LSU2 state), level $|- \rangle$ decouples from the electron reservoir,
staying coupled to the rest of the system through its many-body capacitive coupling to level $|+ \rangle$.
}
\label{fig1}
\end{figure}

\begin{figure}
\centerline{\includegraphics[width=3.65in]{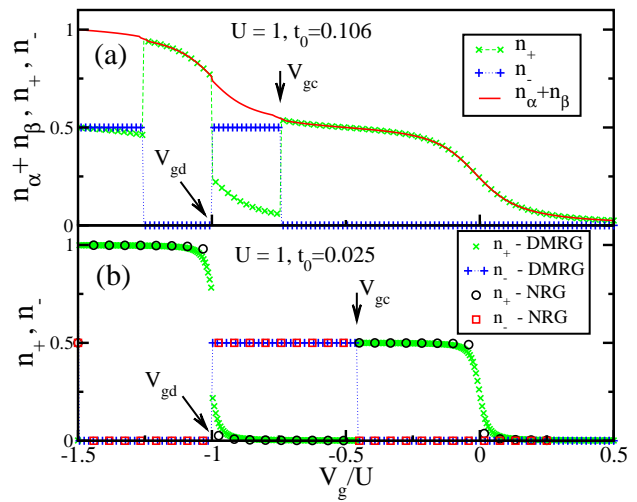}}
\caption{(a) DMRG results for charge variation against gate potential: $n_+(V_g)$, (green) $\times$ symbols,
$n_-(V_g)$, (blue) $+$ symbols, and $n_{\alpha}+n_{\beta}$, (red) solid line (note that $n_{\alpha}=n_{\beta}$).
The charging of the AOs is smooth, while the charging of the MOs has clear discontinuities.
The parameters used are $\phi=\pi/4$, $t_0=0.106$, $U=1$, and $\Gamma=0.045$ (note that
the particle-hole symmetric point of this model is at $V_g=-3U/2$, in addition, the charge is plotted {\it per spin species}).
(b) Comparison between NRG and DMRG $n_{\pm}$ $vs.$ $V_g$ results for the 2LSU2 state
($\phi=\pi/4$, $t_0=0.025$ ($\Gamma=0.0025$), and $U=1$).
NRG: $n_+$ [circles (black)], $n_-$ [squares (red)]. DMRG: $n_+$ [$\times$ (green)], $n_-$ [$+$ (blue)].
Note the very good agreement between the two techniques. DMRG results obtained with 40 sites and maintaining 200 states.
NRG results obtained at the last iteration (N=80), and a discretization parameter
$\Lambda=2.5$ was used, keeping $N_s=1500$ states in each iteration, which allows a larger $N_s$ to account for degeneracies
(all the other NRG calculations in this work were done similarly).
The small differences occur because the DMRG results are not totally converged yet. Note also [as in panel (a)] the interesting
$V_g$-dependent charge oscillations involving the two MOs simultaneously, given that they are capacitively coupled.
See text for definition of $V_{gc}$ and $V_{gd}$.
}
\label{fig2}
\end{figure}

\section{Hamiltonian and Molecular Orbitals}
\subsection{Model Hamiltonian}
The SU(4) Hamiltonian used to model a CNT QD (or a capacitively coupled DQD)\cite{first-exp} coupled to metallic leads is given by
\begin{eqnarray}
 H_{\rm tot} &=& H_{\rm mb} + H_{\rm band}
 + H_{\rm hyb} ,\\
  \label{eq1}
H_{\rm mb}&=&\sum_{\sigma;\lambda=\alpha,\beta}
\left[ {U \over 2} n_{\lambda \sigma} n_{\lambda \bar{\sigma}} + V_g n_{\lambda \sigma}\right] +  \nonumber \\
  && U^{\prime} \sum_{\sigma \sigma^{\prime}} n_{\alpha \sigma} n_{\beta \sigma'} ,\\
  \label{eq2}
 H_{\rm band} &=&  t \sum_{\lambda=\alpha, \beta} 
 \sum_{i=1,\sigma}^\infty 
  (c_{\lambda i\sigma}^+ c_{\lambda i+1\sigma} +  \nonumber \\
  && c_{\lambda -i\sigma}^+ c_{\lambda -i-1\sigma})  
  +\mbox{H.c.} , \\
  \label{eq3}
H_{\rm hyb} &=& \sum_{\lambda,\lambda^{\prime}=\alpha,\beta,i,\sigma} t_{\lambda \lambda^{\prime}}
\left[ d_{\lambda \sigma}^{\dagger} c_{\lambda^{\prime}\pm1\sigma} + \mbox{H.c.} \right] .
  \label{eq4}
\end{eqnarray} 
Our unit of energy is $t$, and all results were obtained for $U=U^{\prime}=t$. Note that this Hamiltonian is 
p-h symmetric for $V_g/U=-1.5$. 
$H_{\rm mb}$ contains all the many-body terms in the system; more specifically, $\lambda=\alpha, \beta$ are
two degenerate orbitals associated to the wrapping mode (clockwise or anticlockwise)
of the electron propagation along the axial direction of the CNT,
while $d_{\alpha \sigma}$ ($d_{\beta \sigma}$) annihilates an electron with spin $\sigma$ in the
$\alpha$ ($\beta$) orbital in the CNT QD and $c_{\lambda i \sigma}$ 
annihilates an electron with spin $\sigma$ in the i-th site of the
$\lambda = \alpha$ or $\beta$ channel for $i>0$ or $i<0$ [right or left leads, respectively, see Fig.~1(a)].
Therefore, each lead has two bands, also named $\alpha$ and $\beta$,
after the notation used for the orbital levels in the QD. Justifications for this assumption
are given elsewhere.\cite{martins1}
It is important to stress that the CNT QD is coupled to two independent metallic leads on each side (at left and right), 
which have the same chiral states as the CNT QD (and therefore the same quantum numbers). 
Equations (3) and (4) describe the leads and their coupling to the 
QD, respectively. More specifically, $t_{\alpha,\alpha} = t_{\beta,\beta} = t^{\prime}$ and 
$t_{\alpha,\beta} = t_{\beta,\alpha} = t^{\prime \prime}$. 
To explore the 2LSU2 regime, we wish to perform numerical calculations for $\nu \lesssim 1$.
Following Ref.~\onlinecite{Aguado}, we will define a ratio
$\phi = \mbox{tan}^{-1}\left(\nu \right)$, with which the hoppings between the QD and leads can be rewritten as
$t^{\prime \prime} = t_0 \sin(\phi)$ and $t^{\prime} = t_0 \cos(\phi)$, where $t_0$ is kept constant,
and $t^{\prime}$ and $t^{\prime \prime}$ are indicated in Fig.~1(a).
Using this definition ensures that a constant coupling $\Gamma=2 \pi \rho_0\left(E_F \right) t_0^2$ 
(where $\rho_0$ is the reservoir's DOS at the Fermi energy) between the QD and the charge reservoirs
is kept constant, while the ratio $\nu$ varies. Note that $\nu=0$ ($t^{\prime \prime}=0$ and $t^{\prime}=t_0$), 
the pure SU(4) state, implies $\phi = 0$,
while $\nu=1.0$ ($t^{\prime}=t^{\prime \prime}=t_0/\sqrt{2}$), the pure 2LSU2 state, implies $\phi = \pi / 4$.
In what follows we will assume, for simplicity, that $U=U^{\prime}$.\cite{note1} 

Figure 1(a) shows a schematic representation of $H_{\rm tot}$, together 
with diagrams indicating the gradual decoupling of bands from the QD [Figs.~1(b) and 1(c)] as the symmetries 
of the system are exploited, finalizing 
with the construction of the MOs $|+ \rangle$ and $|- \rangle$. A brief description of these transformations 
is given below and full details are provided in Appendix A.

\subsection{Molecular orbitals: band decoupling}

Given the left-right symmetry of the system, two bands decouple from the QDs, as indicated in 
Fig.~1(b), through a symmetric-antisymmetric combination of creation-annihilation 
operators from left and right leads (without mixing $\alpha$ and $\beta$ states). 
The resulting system [see Fig.~1(b)] still has one more symmetry to be exploited. If one performs 
a symmetric-antisymmetric linear combination of $\alpha^{even}$ and $\beta^{even}$ creation-annihilation operators 
acting on sites situated at the same distance from the QD [see Fig.~1(b)] {\it and} a bonding-antibonding 
linear combination of the QD levels $|\alpha \rangle$ and $|\beta \rangle$, 
one obtains the MOs $|+ \rangle$ and $|- \rangle$, which couple to symmetric and 
antisymmetric bands, respectively, as shown in Fig.~1(c), through hopping matrix elements 
\begin{eqnarray}
t_{+} &=& \sqrt{2}(t^{\prime} + t^{\prime \prime}) \label{eq5} \\
t_{-} &=& \sqrt{2}(t^{\prime} - t^{\prime \prime})
  \label{eq6}
\end{eqnarray}

It is now clear that, {\it at} the 2LSU2 point (where $\nu=1$, i.e., $t^{\prime}=t^{\prime \prime}$, 
and therefore $t_{-}=0$), a third band decouples from the system, leaving MO $|- \rangle$ 
decoupled from any electron reservoir, although still capacitively coupled to MO $|+ \rangle$.
This immediately answers some of the questions raised above. First, in the limit $\nu \rightarrow 1$, 
the charge transport changes from double- to single-channel. Therefore, the maximum conductance 
will change from $2G_0$ to $G_0$. Indeed, in the vicinity of the 2LSU2 point, for $\nu \lesssim 1.0$, 
narrow peaks with height $2G_0$ are expected around certain values of gate potential. The closer one gets to the 
2LSU2 point, the narrower these peaks become, their width vanishing at the 2LSU2 point, where 
the maximum of the conductance will be $G_0$.\cite{martins1,martins2} Second, at the 2LSU2 state, 
MO $|- \rangle$, being decoupled from the charge reservoir, will have zero width,  
i.e., $\Gamma_-$ vanishes (for a definition of $\Gamma_-$, see below and Appendix A). 
As the gate potential varies, and after the level $|- \rangle$  
has crossed the Fermi energy, it will be abruptly charged, changing the ground state 
of the system (recall that MO $|- \rangle$ has a many-body interaction with the rest of the system), 
and potentially changing its transport properties discontinuously. 
One should expect that the Kondo peak in the LDOS, when the system is in the SU(4) state, 
should differ considerably from the Kondo peak occurring in the 2LSU2 state, given the difference 
in the number of charge transport channels between the two states.
The very abrupt change in the slave boson's LDOS (seen in Fig.~14 of reference \onlinecite{Aguado}) 
is probably linked to this change from double- to single-channel charge transport. If 
the change in the Kondo resonance is discontinuous or not (at $\phi=\pi/4$) is still a point to 
be further investigated. 

For completeness, we present here $H_{\rm mb}$ written in the new basis, assuming that $U=U^{\prime}$ 
(in Appendix A, a complete expression, for the case $U \neq U^{\prime}$ is presented), 
\begin{eqnarray}
H_{\rm mb}&=&\frac{U+U^{\prime}}{4}\left[\sum_{\lambda,\sigma} n_{\lambda \sigma} n_{\lambda \bar{\sigma}} 
+ \sum_{\lambda \neq \lambda^{\prime},\sigma} n_{\lambda \sigma} n_{\lambda^{\prime} \bar{\sigma}}\right] + \nonumber \\
 && \frac{U^{\prime}}{2} \sum_{\lambda \neq \lambda^{\prime},\sigma} n_{\lambda\sigma} n_{\lambda^{\prime}\sigma}
\end{eqnarray}
where $\lambda=\pm$, $\sigma=\uparrow$ or $\downarrow$, and although our calculations are for $U=U^{\prime}$, 
both are indicated, with their respective terms.
It is also useful to define the broadening of the one-body AOs $|\alpha,\beta \rangle$, as well as for 
the MOs  $|\pm\rangle$ (coming from their coupling to the leads). For the AOs, we have
\begin{eqnarray}
\Gamma &=& \Gamma_\alpha + \Gamma_\beta , \\
\Gamma_{\alpha/\beta} &=& 2\pi (t'^2+t''^2) \rho_0(E_F) = 2\pi t_0^2 \rho_0(E_F) \label{Aeq13b},
\end{eqnarray}
where the factor 2 in Eq.~\ref{Aeq13b} takes in account the left {\it and} right leads [see Fig.1~(a)], 
and $\rho_0(E_F)$ is the LDOS at the first site of the leads (here considered the same for the $\alpha$ and $\beta$ channels), and $\Gamma$ 
is the total coupling between the interacting region and the leads. 
For the MOs we have
\begin{eqnarray}
\Gamma_+ &=& \pi t_+^2 \rho_0(E_F) ,\\
\Gamma_- &=& \pi t_-^2 \rho_0(E_F) .
\end{eqnarray}
Note that, as required, $\Gamma = \Gamma_+ + \Gamma_-$ (for more details, see Appendix A).

The MO $|-\rangle$, in the 2LSU2 regime ($\nu = 1$), is an electronic {\it dark state} that cannot 
interact directly with the conduction electrons. Such state has been extensively 
studied in different systems, namely, as Dicke effect \cite{Pedro1, Pedro2, Marcelo1} or 
bound states in the continuum (BICs), \cite{Guevara2, solis} which can produce Fano 
resonances in the conductance. \cite{Guevara1} The original Dicke effect 
in quantum optics \cite{dicke} takes place through 
spontaneous emission, by two closely spaced atoms (of the same species), which emit a photon into
a common environment. 

\begin{figure}
\centerline{\includegraphics[width=3.65in]{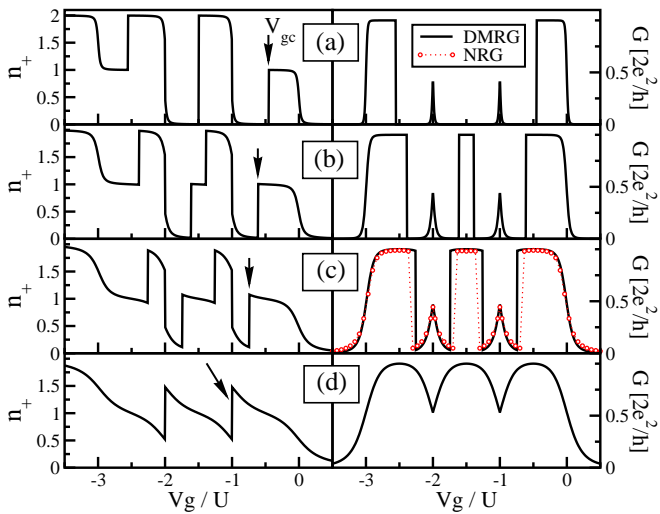}}
\caption{ Left panels: DMRG results similar to the ones in Fig.~2, but now for varying $t_0$ ($0.025$, $0.05$, $0.106$,
and $0.175$, from top to bottom). The corresponding values for $\Gamma$ are $0.0025$, $0.01$, $0.045$,
and $0.1225$. $U=1.0$ for all panels. For the sake of clarity of presentation,
just the $n_+$ results are shown. Note that the value of $V_{gc}$ [gate potential for charging (discharging) of $|- \rangle$ ($|+ \rangle)$] is
strongly dependent on $\Gamma$, while $V_{gd}$ (gate potential for discharging (recharging) of $|- \rangle$ ($|+ \rangle)$] is not
(its value is practically fixed at $V_g/U = -1.0$).
Right panels: Solid (black) lines: conductance obtained through the Friedel sum rule applied to the corresponding charges in
the left panels. (Red) Dots in panel (c): NRG results for the conductance. Notice the good agreement between the two techniques.
The narrow peaks in the 3 top right panels, located at $V_g/U=-1.0$, are briefly discussed in the text.
OBS.: The bottom right panel should be compared to Fig.~4 in Ref.~\onlinecite{martins1}.
The fact that only one NRG result is shown in the right-side panels is explained in the text.
}
\label{fig3}
\end{figure}

\begin{figure}
\centerline{\includegraphics[width=3.0in]{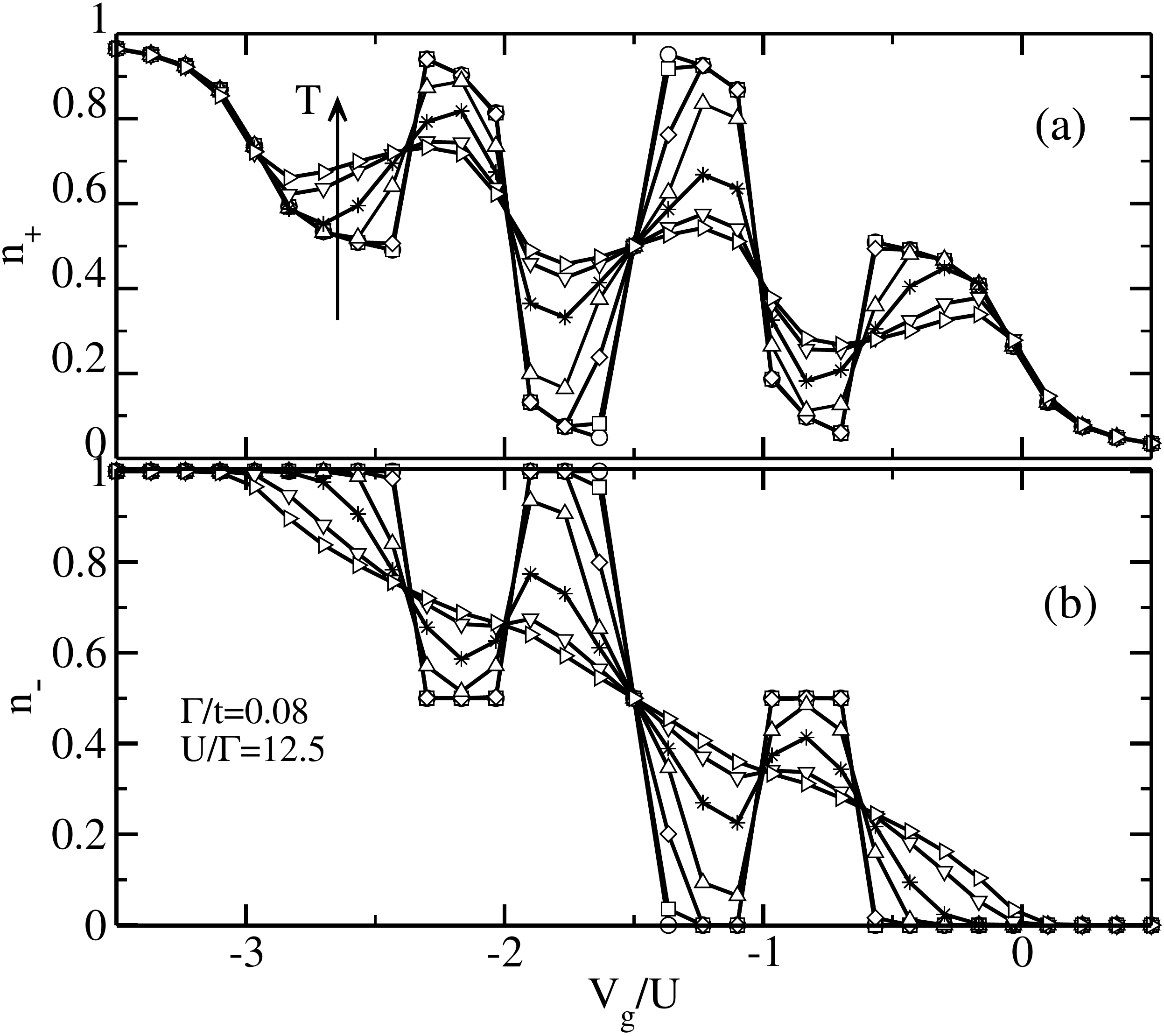}}
\caption{Temperature (in units of $U$) dependent NRG results for $n_{\pm}$ $vs.$ $V_g$. Maximum temperature
(right pointing triangles) is $T=3.118 \times 10^{-2}$
[minimum temperature is zero (circles)]. The arrow in panel (a) indicates direction of temperature increase. Note that
the MO $V_g$-dependent charge oscillations have been strongly suppressed at $T=7.888 \times 10^{-3}$ (stars),
a value that is of the same order of magnitude
as the Kondo temperature of a corresponding (same parameters) single impurity problem. This suggests that
the MO charge oscillations arise from a low-energy effect, possibly related to the Kondo effect.
Parameters are indicated in the bottom panel.
}
\label{fig4}
\end{figure}

\section{Charge oscillation of molecular orbitals: DMRG and NRG results}

\subsection{DMRG and NRG (zero and finite-temperature) results}

Figure 2(a) contrasts DMRG results at the 2LSU2 state ($\phi=\pi/4$) obtained for $n_{\pm}$ $vs.$ $V_g$ and $n_{\alpha,\beta}$ $vs.$ $V_g$, 
emphasizing the difference between them: the AOs present a smooth charging, as a function of $V_g$, 
and, obviously, $n_{\alpha}\left(V_g \right) = n_{\beta}\left(V_g \right)$, while $n_{\pm}\left(V_g \right)$ 
show clear discontinuities and very diverse behavior. Note that we plot only up to the 
p-h symmetric point ($V_g/U=-1.5$), and the data has to be multiplied by $2$ to reproduce the total charge 
in the QD, i.e., we are plotting charge per spin species, as the system is SU(2) invariant. 
Figure 2(b) shows a comparison of $n_{\pm}$ $vs.$ $V_g$ between NRG and 
DMRG, at zero-temperature. The agreement between the two techniques is quite good, as can be easily seen. 
The hopping and many-body parameters used are indicated in the figure and in the caption, and $\phi=\pi/4$ 
for both panels (2LSU2 {\it state}). As mentioned above, at the 2LSU2 state, level $|- \rangle$ is completely decoupled  
from the electron reservoir and will obviously be abruptly charged for some $V_g$ value, and, 
given that it is capacitively coupled to level $|+ \rangle$, its occupancy will also change abruptly. This 
is clearly reflected in the results for $n_{\pm}$ in Figs.~2(a) and (b). Note that the results in 
Fig.~2(a) are for a smaller value of $U/\Gamma$ than those in Fig.~2(b) (the second is deeper into the 
Kondo regime), although this discontinuity in the charge occupancy of the MOs still occurs for both cases.
Note that the log-discretization of the semi-infinite chain used by NRG is the one described in 
Ref.~\onlinecite{jayaprakash}.

The questions to be answered here are the following: (i) what 
determines the value of $V_g$ for which the charging (discharging) and discharging (recharging) of level 
$|- \rangle$ ($|+ \rangle$) occurs (indicated in Fig.~2(a) with arrows and denoted $V_{gc}$ and 
$V_{gd}$)? (ii) How this value depends on $\Gamma$? 
(iii) And finally, how robust is this phenomenon? I.e., how far from the {\it strict} 2LSU2 regime 
can we go (i.e., for $\phi<\pi/4$) and still observe this gate potential dependent MO charge oscillation effect? Why this 
charge oscillation effect is important will be apparent soon (see Fig.~6). 

\begin{figure}
\centerline{\includegraphics[width=3.5in]{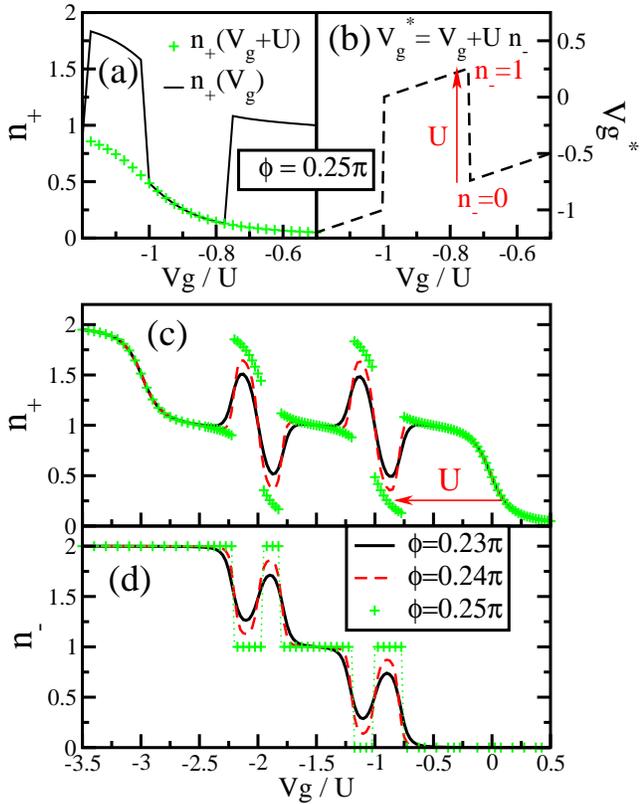}}
\caption{(a) $n_+$ $vs.$ $V_g$ DMRG results ($U=1.0$, $\Gamma=0.045$, and $\phi=\pi/4$) 
comparing $n_+ \left(V_g \right)$, solid black line,
with $n_+ \left(V_g + U \right)$ [(green) $+$ symbols] for $V_g$ values around $-0.9U$.
The perfect agreement between the two curves, in the region between discontinuities, indicates the
simple fact that the charging of the decoupled MO level $|- \rangle$
acts as an extra electrostatic potential, moving up by $U$ the effective potential
seen by MO level $|+ \rangle$.
This is schematically shown in (b), where the effective potential is plotted against $V_g$ 
and the (red) vertical arrow indicates the charging of MO $|- \rangle$ and the 
change by U of the effective potential seen by MO $|+ \rangle$ (the discharging of MO $|- \rangle$ 
occurs in the other discontinuity). 
(c) and (d):  DMRG results for $n_{\pm}$ $vs.$ $V_g$ for
three different values of $\phi/\pi$ ($0.25$, $0.24$, and $0.23$). This shows that
the $V_g$-dependent charge oscillations seen {\it at} $\phi/\pi=0.25$ survive away from the
2LSU2 point. The charge oscillations are still noticeable down to
$\phi/\pi \approx 0.2$ (not shown). The parameters are $t_0 = 0.106$, $\Gamma = 0.045$, and $U=1.0$.
The DMRG calculation was done for a cluster with 80 sites and 200 states were kept, 
and for $\phi < 0.25$ only the left-right transformation was used [see Fig.~1(b)] 
The black horizontal arrow in panel (c), when applied to the (green) $+$ symbols curve, 
schematically shows the displacement presented in panel (a).
}
\label{fig5}
\end{figure}

Figure 3, showing DMRG results for different values of $U/\Gamma$, answers the second question. I.e., 
how does the charging of MO level $|- \rangle$ depends on $\Gamma$ (for fixed $U$)?
In it, on the left side panels, one sees that $V_{gc}$ depends on $\Gamma$ much more strongly than $V_{gd}$, as the value of 
$V_{gd}$ barely changes with $\Gamma$, while $V_{gc}$ decreases as $\Gamma$ increases, until 
$V_{gc}=V_{gd}$ (a full explanation of this effect will be given 
in section V). To underscore the fact that for $\phi=\pi/4$ charge transport occurs {\it only} through 
MO $|+ \rangle$, the right panels in Fig.~3 show the conductance results obtained by using the Friedel sum rule (solid curves) 
$G=G_0 \sin^2(\pi n_+/2)$, where the results used for $n_+(V_g)$ are those obtained by DMRG (left panels). 
The dotted curve in panel (c) shows the conductance results obtained by NRG (through the calculation of the dressed 
Green's function). The good agreement between both techniques indicates 
that, as expected, the conductance is all through the molecular orbital $|+ \rangle$. The origin of the 
sharp peak at $V_g/U=-1.0$ will be explained in section IV, where the LDECA results for conductance are presented. 
Note that the lack of NRG results for the other right-side panels is due to convergence issues. \cite{note6}

\begin{figure}
\centerline{\includegraphics[width=3.65in]{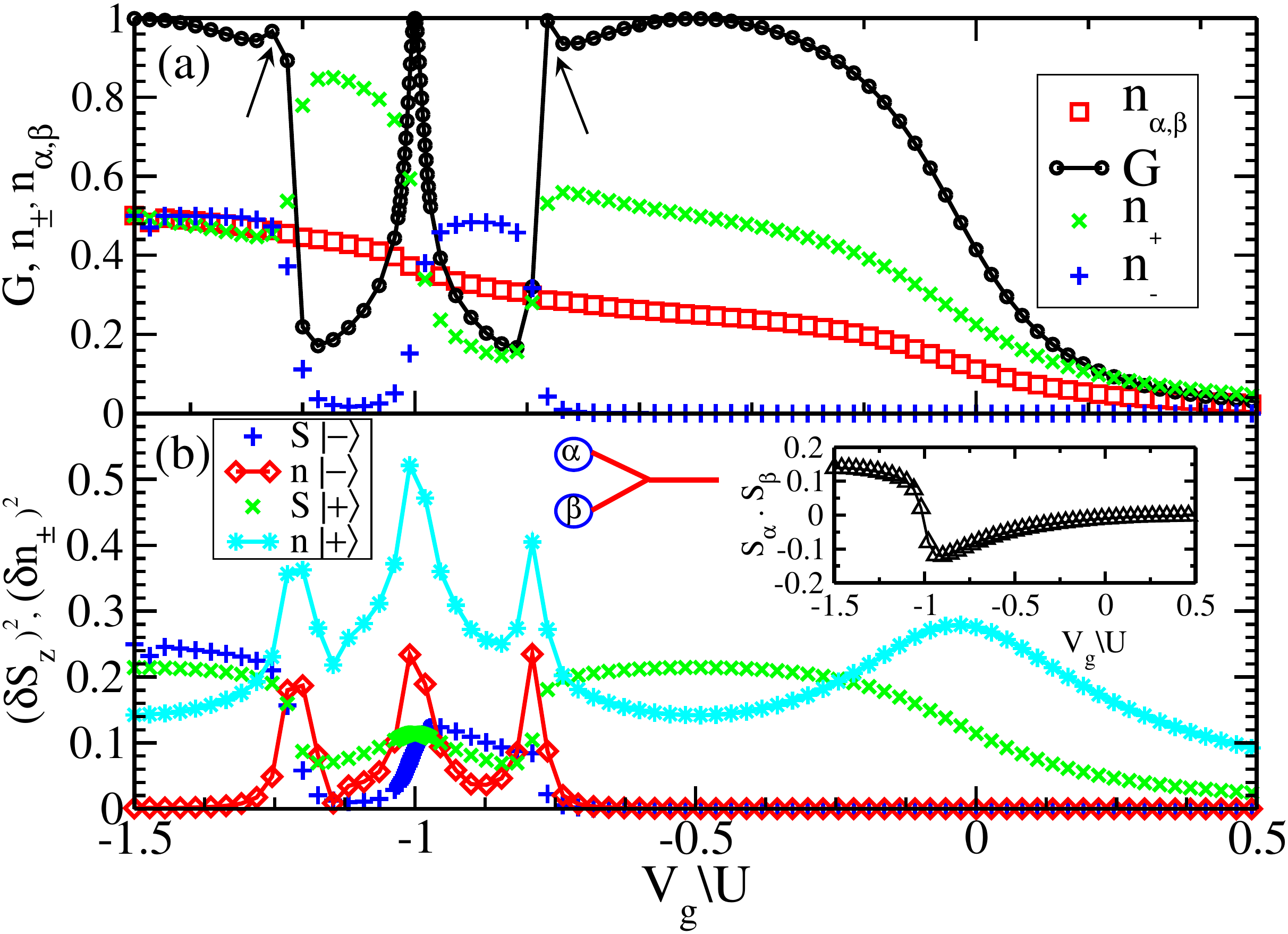}}
\caption{(a) LDECA results ($t_0=0.15$, $\Gamma=0.09$, $U=1.0$, $\lambda=7$, and $\phi=0.242$) for conductance $G$ [(black) circles], 
charging of the MO levels $n_{\pm}$ ($\times$ (green) and $+$ (blue) symbols 
for $n_+$ and $n_-$, respectively), and charging of the AO levels $n_{\alpha,\beta}$ 
(square (red) symbols). The mere comparison of $G$ with $n_{\pm}$
and with $n_{\alpha,\beta}$ indicates that the conductance variation with $V_g$ is intimately correlated with that 
of $n_{\pm}$, i.e., understanding the $V_g$-dependent charge oscillations of the MOs $|\pm \rangle$ 
is crucial to the understanding of the charge transport properties of the system. 
In addition, by comparison with Figs.~5(c) and (d),
one notices the qualitative agreement between the DMRG and LDECA results for
$n_{\pm}$ as a function of $V_g$. (b) LDECA results for spin and charge fluctuations for both MOs. 
$\times$ symbols (green) and $+$ symbols (blue) indicate $\left(\delta S_z \right)^2$ for MO $|+ \rangle$ 
and $|- \rangle$, respectively; diamonds (red) and $\ast$ (cyan) indicate $\left(\delta n \right)^2$ 
for MO $|- \rangle$ and $|+ \rangle$, respectively. 
A plateau in $\left(\delta S_z \right)^2$, accompanied by simultaneous suppression of charge fluctuations, 
indicates a level participating in a Kondo effect. 
It is interesting to note that, according to these results, MO level $|- \rangle$, although very weakly coupled
to the charge reservoir, and therefore still undergoing strong $V_g$-dependent charge oscillations
(notice that $\phi \sim \pi/4$), seems to participate in the Kondo
effect, mainly close to the p-h symmetric point, where its spin fluctuations are higher than those for MO $|+ \rangle$. 
The inset shows the results for spin-spin correlations between the AO levels $|\alpha \rangle$ 
and $|\beta \rangle$ (vertical triangles). Note that close to the p-h symmetric point they are ferromagnetically correlated. 
The cartoon at the left side of the inset explains why (see text and Reference~\onlinecite{ferro-PRL}).
The conductance structures indicated by small arrows in panel (a) are discussed in the text. 
Note that the charge fluctuation for MO $|+ \rangle$ [$\ast$ (cyan)] was divided by $2$ to fit better in panel (b). 
}
\label{fig6}
\end{figure}

To help answer the other questions, it is important to 
have an idea of the order of magnitude of the energy scale of the processes that determine this 
gate potential dependent charge oscillation. The temperature (in units of $U$) dependent $n_{\pm}$ $vs.$ $V_g$ NRG results in Fig.~4 
indicate that these $V_g$-dependent charge oscillations have been strongly suppressed at a temperature which 
is a very small fraction of $U$ (see temperature values in the caption). Indeed, as the charge transport occurs through 
level $|+ \rangle$, as shown by the description of the DMRG results presented in Fig.~3, 
the Kondo temperature of the system is in reality that of a single QD with the same $\Gamma$ and $U$ as 
that of level $|+ \rangle$. This Kondo temperature,  
calculated using NRG, is of the same order of magnitude as the temperature for which 
the $V_g$-dependent charge oscillations have been strongly suppressed ($T \approx 7.888 \times 10^{-3}$, $\ast$ curve). This indicates that, 
as initially thought, these charge oscillations are governed by a many-body process related to the Kondo effect, 
or some other many-body effect with a similar characteristic energy (as shown in Appendix B, 
the non-interacting model, $U^{\prime}=U=0$, does not show these $V_g$-dependent charge oscillations).

\subsection{The effective gate potential model ($\phi=\pi/4$)}

From the point of view of level $|+ \rangle$, level $|- \rangle$, when charged, 
acts as an extra source of electrostatic energy, besides the one provided by the 
back-gate ($V_g$). The particularities here are that, {\it at} the 2LSU2 point ($\phi=\pi/4$), 
level $|- \rangle$, being decoupled from the electron reservoir, can only accommodate 
an integer number of electrons ($n_- = 0$, $1$, or $2$). Taking this into account, 
one can define an effective gate potential as $V_g^{\ast}=V_g + Un_-$.
From this point of view, it is then easy to understand the effect of 
the charging and discharging of level $|- \rangle$: it just modifies 
the {\it effective} gate potential seen by level $|+ \rangle$. To emphasize that this is 
indeed the only effect in play for the charging of level $|+ \rangle$ at the 2LSU2 state, 
we show in Fig.~5(a) the curves for $n_+(V_g)$ and $n_+(V_g+U)$ for 
$V_g/U$ values around $-0.9$ [for $U/\Gamma=22.0$, corresponding to 
Fig.~3(c)]. The horizontal arrow in Fig.~5(c) schematically shows 
the `superposition' of the two curves shown in Fig.~5(a). 
The matching of the two curves in Fig.~5(a), for $n_+(V_g)$ [solid (black)] and $n_+(V_g+U)$ [(green) $+$ symbols], 
in the region between discontinuities, is perfect, confirming the effective potential 
idea described above: the discontinuous charging of the MO $|- \rangle$ changes by U the 
gate potential seen by MO $|+ \rangle$, discharging it by 1 electron 
(as the total charge in the system is kept constant), and, more importantly, 
as $V_g$ decreases further below $V_{gc}$, the {\it recharging} of MO $|+ \rangle$ 
occurs exactly the same way [see Fig.~5(a)] as when MO $|- \rangle$ was empty, at $V_g=V_{gc}+U$. 
As a reference, the effective gate potential $V_g^{\ast}$, as a function of $V_g$, is shown 
in Fig.~5(b). The question still to be answered is what physical process determines $V_{gc}(\Gamma)$, 
i.e., the dependence of the value of $V_{gc}$ with the coupling to the leads $\Gamma$ (as depicted in 
the left-side panels of Fig.~3). Before 
answering this question, let us concentrate on the other two 
pending issues: how robust is this gate-dependent charge oscillation away from the exact 
2LSU2 point, and what is its effect over the charge transport, i.e., the conductance of the system.

\section{DMRG and LDECA results away from the 2LSU2 point}

Figures 5(c) and (d) show DMRG $n_{\pm}$ $vs.$ $V_g$ results for three different 
values of $\phi/\pi$ ($0.25$, $0.24$, and $0.23$). As expected, 
the discontinuities are smoothened out, but the charge oscillations 
survive. Additional calculations (not shown) indicate that the oscillations are 
still noticeable for $\phi/\pi \approx 0.2$ ($t^{\prime \prime}/t^{\prime} \approx 0.73$). 
Once $\phi < \pi/4$, the transport problem becomes a 
double channel one, making its solution through NRG much more 
costly in computational terms. To circumvent this, here we use the 
LDECA method \cite{ldeca} to present results for $\phi < \pi/4$. Fig.~6(a) 
shows results for conductance $G$ (* (black) symbols), and 
$n_{\pm}$ $vs.$ $V_g$ for $\phi/\pi = 0.242$ 
($\times$ (green) and $+$ (blue) symbols for $n_+$ and $n_-$, respectively). 
The (red) square symbols show the charging of the AO levels $|\alpha,~\beta \rangle$. A comparison of 
the conductance curve with the curves for the charging of the MO levels and the AO levels indicates 
that the charging of the AO levels provides very little information about the sharp features observed 
in the conductance. These features are obviously more correlated to the charging of the MOs. As 
$\phi \approx \pi/4$, we are still too close to the single-channel regime, and the conductance 
is still too close to those obtained in Fig.~3 (right-side panels). 
The charging results for the MO levels 
are in good qualitative (and even quantitative) agreement 
with those for DMRG [as shown in Figs.~5(c) and (d)]. 

It is interesting to contrast the very sharp peak in the conductance, around the point 
where level $|- \rangle$ is discharged, with the ones shown in the right-side panels 
in Fig.~3. In Fig.~5(a), the peak reaches one quantum of conductance ($G_0$), while in Fig.~3 it reaches just 
$\approx G_0/2$). Although the convergence of LDECA results as $\phi \to \pi/4$ becomes 
increasingly difficult, the peak at $V_g=-U$, for $\phi < \pi/4$, has always a 
maximum of $G_0$. We speculate that the difference with the $\phi = \pi/4$ results (Fig.~3) stems from the number of channels. Although 
$\Gamma_-$ is very small when $\phi \approx \pi/4$, there {\it are} two channels available for charge 
transport (contrary to the situation at $\phi = \pi/4$). Notice in Fig.5(a) that the discharging of MO 
$|- \rangle$ (at $V_g/U \approx -1$) occurs over a window of $V_g$ (with width $\Gamma_-$), indicating that level 
$|- \rangle$ is in resonance with the band, therefore participating in charge transport. 
LDECA results for smaller
values of $\phi$ (down to $\phi/\pi \sim 0.2$) indicate a smooth evolution
of the conductance, until values of $G$ start to show features
above $G_0$, once the double channel transport
takes hold (and the conductance reaches $2G_0$ for some narrow regions
of $V_g$).\cite{note2}.
In addition, a comparison of the right side tail of the narrow peak, with the 
corresponding right tail of the broad plateau (approximately centered 
around $V_g/U=-U/2$) shows that the effective potential idea is still 
helpful in interpreting the results (at least for this small deviation from $\phi=\pi/4$). 
Note that $n_-$ (+ (blue) symbols) is still very approximately quantized. 
Finally, it is interesting to point out that the small upticks in the conductance, 
before the sharp drop, and after the sharp rise (indicated with arrows) 
have been seen in functional Renormalization Group (fRG) calculations, for a similar 
model, by C. Karrasch {\it et al.} [see their Fig.~5(a)].\cite{meden} 

In Fig.~6(b), results for spin and charge fluctuations for the MOs are shown. As 
expected, the rightmost conductance plateau in panel (a) is associated 
to a plateau in the spin fluctuations of MO level $| + \rangle$ ($\times$ (green) symbols), 
and a suppression of its charge fluctuations (* (cyan) symbols), which are 
fingerprints of the Kondo effect. Taking in account the comment above, that the 
effective gate potential model is still useful, the narrow peak around $V_g/U=-1.0$ 
can then be interpreted as a Kondo state that was `skipped' due to the 
second charge oscillation: indeed, when MO $|- \rangle$ discharges approximately 1 electron, 
the effective potential seen by MO $|+ \rangle$ abruptly decreases by $\sim U$, 
moving the system from the high $V_g$ side of the Intermediate Valence regime 
to its low $V_g$ side, completely skipping the Kondo regime (note the enhancement 
of charge fluctuations for MO $|+ \rangle$ around $V_g/U=-1.0$, and a much smaller 
value for its spin fluctuations, as compared to the right-side maximum, 
indicating that the charge transport process here is more akin to sequential tunneling)\cite{note3}

Around the p-h symmetric point, a Kondo effect with peculiar characteristics arises, 
where {\it both} MOs participate. This can be confirmed by checking the spin fluctuations 
of MOs $|+ \rangle$ {\it and} $|- \rangle$. In reality, the spin fluctuations of MO $|- \rangle$ 
($+$ (blue) symbols) are (surprisingly!) more enhanced 
than those of MO $|+ \rangle$. Given the capacitive coupling between the two MOs, 
the participation of level $|+ \rangle$ in a Kondo effect drives level $|- \rangle$ into also 
becoming an active degree of freedom in the Kondo effect. A somewhat different view of this Kondo effect 
can be taken if one returns to the AO basis and calculates the spin correlations 
between AOs $|\alpha \rangle$ and $|\beta \rangle$. This is shown in the inset in 
Fig.~6(b): the two AO states, close to the p-h symmetric point develop a {\it ferromagnetic} 
correlation. This can be readily understood by referring to Fig.~1(b): instead of performing 
the symmetric/antisymmetric {\it and} the bonding/anti-bonding transformations that lead 
to Fig.~1(c), one should skip the second transformation (keeping the AOs). This leads 
to (for $\nu=1$) the schematics shown at the left side of the inset in Fig.~6(b), which is well known to 
lead to ferromagnetic correlations between the AOs, as remarked by Martins {\it et al.}\cite{ferro-PRL} 
and confirmed in other publications thereafter. \cite{ferro-JPCM} 

\begin{figure}
\centerline{\includegraphics[width=3.6in]{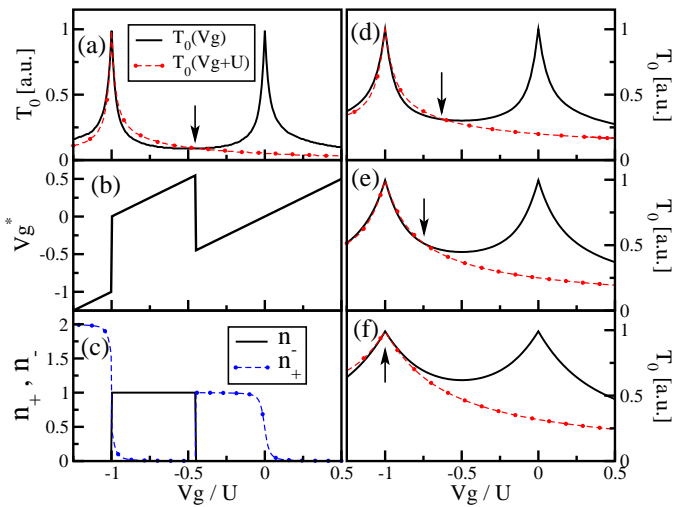}}
\caption{(a) Comparison of $T_0(V_g)$ (solid (black) curve) with $T_0(V_g+U)$ (dotted (red) curve)
for a large window of $V_g$. Notice the crossing point for the two curves (indicated by an arrow).
As can clearly be seen, this is the point for which the gain in energy ($T_0$) by
`moving up' the effective gate potential of the system by U, and going back to the Intermediate Valence regime,
is larger than the gain in energy provided by `staying' in the Kondo state.
It can be clearly seen in panels (b) and (c), which show the effective gate potential
and corresponding results for $n_{\pm}(V_g)$, respectively, that the crossing indicated
in panel (a) exactly agrees with $V_{gc}$, i.e., the point where $n_-$ is charged
and $n_+$ is discharged. In addition, the second charge oscillation (at $V_{gd}/U=-1$)
occurs at the point where $T_0(V_g+U)$ crosses $T_0(V_g)$ for the second time, but now
passing {\it under it}. In this case, it is energetically favorable for the system
to `move down' the effective potential, i.e.,
when the system starts to enter the Kondo regime again, it is energetically advantageous
for the system to `skip' the whole Kondo
regime and go straight to the second Intermediate Valence regime. This is accomplished by MO level
$| - \rangle$ being discharged and level $| + \rangle$ being recharged. Parameters
used in panels (a) to (c) are the same as the ones used in Fig.~3(a).
Panels
(d) to (f) show similar results as in panel (a), but for progressively larger
values of $\Gamma$ (the same parameters as the ones used in panels (b) to (d)
in Fig.~3). For all the cases analyzed, the crossing of $T_0(V_g)$ and $T_0(V_g+U)$
agrees with $V_{gc}$.
}
\label{fig7}
\end{figure}

\begin{figure}
\centerline{\includegraphics[width=3.5in]{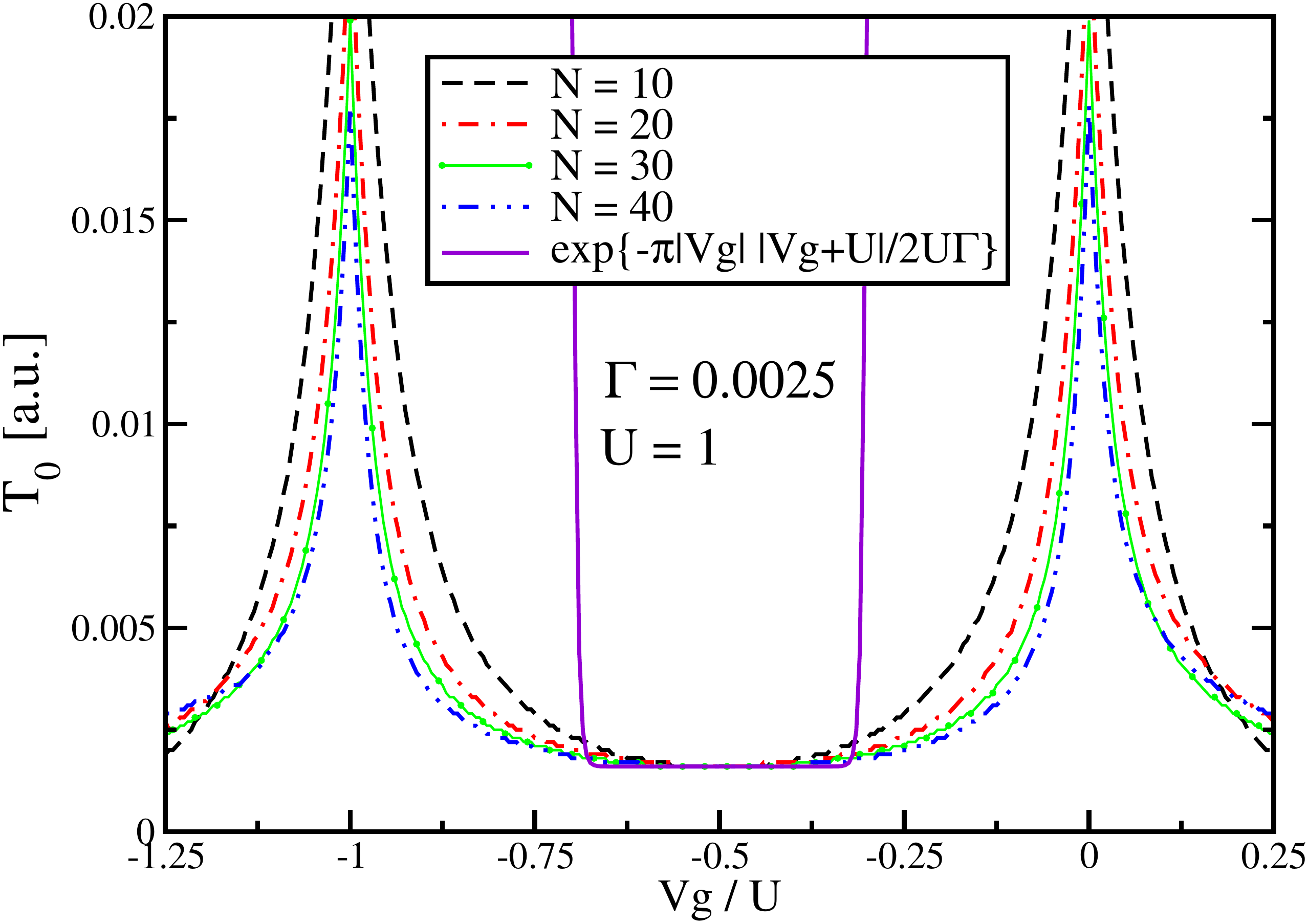}}
\caption{ Comparison of the DMRG finite size scaling calculation of $T_0$ 
(see text for its definition), with the Haldane expression for $T_K$, 
Eq.~(12). Curves for $N=10$, $20$, and $30$, were translated to match 
the $N=40$ curve at $V_g/U=-0.5$. The same translation was applied to 
Haldane's curve. It is clear that, in the region of $V_g$ where the Haldane expression applies (before 
it diverges), the DMRG results for larger clusters 
gradually approach Haldane's results. Parameters are the same as the ones 
used in Fig.~3(a).
}
\label{fig8}
\end{figure}

\section{Competition between Kondo and Intermediate Valence effects ($\phi=\pi/4$)}

The origin of the dependence of $V_{gc}$ with the coupling to the leads ($\Gamma$), 
as shown in the left panels of Fig.~3, still has to be understood. In Fig.~4 we have 
shown that the $V_g$-dependent charge oscillations are related to a low energy scale 
phenomenon. Another important point that has to be remarked, is that the 
very dependence of $V_{gc}$ with $\Gamma$ clearly indicates that the MO level 
$|- \rangle$ does not get charged the moment it crosses the Fermi energy, as 
$V_g$ varies. This conclusion is obvious from the fact that the position of the 
Fermi energy is the same for all left-side panels in Fig.~3, and still, 
the level $|- \rangle$ gets charged at different gate potential values. This is a 
consequence of MO level $|- \rangle$ being 
strongly correlated to the whole system, including the electron reservoir, through 
MO $|+ \rangle$. Bearing in mind now that the charging of level $|- \rangle$ 
{\it at a certain $V_g$ value} results 
in `moving back' level $|+ \rangle$ from a Kondo state to an Intermediate Valence 
state by increasing its effective gate potential by U, one may assume that it is 
energetically more favorable for the system {\it at that 
$V_g$ value} to have MO level $|+ \rangle$ in the Intermediate Valence state than in the Kondo state. By using DMRG to calculate 
the ground state energy of a {\it Single Impurity Anderson Model} (SIAM) as a function of $V_g$, this intuitive idea can be 
quantitatively probed. To achieve that, one just has to remove from the SIAM ground state energy, calculated with DMRG, 
its single-electron contributions, i.e., $H_{\rm band}$, the gate potential energy, and the {\it local} 
correlation energy $U \left( n_{\downarrow}~n_{\uparrow} \right)$, thus 
defining the following energy, here denoted as $T_0 \left(V_g \right)$:
\begin{eqnarray}
T_0(V_g) = E_0(\Gamma=0,V_g) - E_0(\Gamma,V_g),
\end{eqnarray}
where $E_0(\Gamma,V_g)$ indicates the ground state energy (calculated with DMRG) for a SIAM 
with parameters $U=1$, and hopping to the band equal to $t_+=2t_0$ [for $t_0=0.025$, corresponding to Fig.~3(a)], 
while $E_0(\Gamma=0,V_g)$ is the corresponding ground state energy for the same model and same parameters, but now with the impurity 
disconnected from the band. Note that, as $T_0 \left( V_g \right) > 0$, it represents the gain in energy,  
by the system, when the many-body correlations are extended into the leads (be it by the formation of the Kondo state, 
or any other strongly correlated ground state). \cite{note5} 
We find out, as shown in Fig.~7, that this energy $T_0\left( V_g \right)$ explains 
the dependence of $V_{gc}$ with $\Gamma$.

In Fig.~7(a), the energy $T_0$ is plotted as a function of $V_g$, and, 
as done in Fig.~5(a), we compare $T_0(V_g)$ (solid (black) curve) with $T_0(V_g+U)$ (dot-dashed (red) curve). 
The `horizontal displacement' of $T_0$ by $U$ (dot-dashed (red) curve) mimics the effect over MO $|+ \rangle$ (the {\it single impurity} 
in the SIAM) of the charging of MO $|- \rangle$. 
The arrow in Fig.7(a) indicates the crossing point of the two curves, i.e., the point for which 
the {\it gain in many-body energy $T_0$}, by `moving up' the gate potential of the system by U, and 
going back to the Intermediate Valence regime (dot-dashed (red) curve), 
is larger than the gain in energy provided by `staying' in the Kondo state [(black) solid curve].
What affords the double-MO system (with MO levels $|+ \rangle$ and $|- \rangle$) the 
ability to put in direct competition two states relatively 
distant in the phase diagram (the center of the CB valley and the initial ramp up of the CB peak) is the 
presence of the zero-width strongly correlated MO level $| - \rangle$ that can be charged abruptly 
and can only contain an integer number of electrons (for $\phi=\pi/4$, or approximately integer 
for $\phi \lesssim \pi/4$).
Panels (b) and (c) in Fig.~7 show the effective gate potential (as a function of $V_g$) 
and the corresponding results for $n_{\pm}(V_g)$, respectively, for the same parameters 
used to calculate $T_0$ ($U=1.0$, $t_0=0.025$) for the SIAM in panel (a), 
but for the model containing both MO states. The crossing indicated
in panel (a) (vertical arrow) agrees {\it exactly} with $V_{gc}$, i.e., the point where $n_-$ is charged
and $n_+$ is discharged. In addition, the second charge oscillation (at $V_{gd}/U=-1$)
occurs at the point where $T_0(V_g+U)$ crosses $T_0(V_g)$ for the second time, i.e.,
when MO level $|+ \rangle$ starts to enter the Kondo effect again (dot-dashed (red) curve in 
panel (a), around $v_g/U=-1$), it is energetically advantageous
for it to `move down' (by decreasing the effective potential by $U$, as $n_-$ changes from $1$ to $0$), thus `skipping' the whole Kondo
regime and going straight to the second Intermediate Valence regime. This is accomplished by MO level
$| - \rangle$ being discharged and level $| + \rangle$ being recharged (and this `skipping' of the 
Kondo regime results in the narrow peak seen in the conductance in Fig.~5(a), centered around $V_g/U=-1.0$). Panels
(d) to (f) show similar results as in panel (a), but for progressively larger
values of $\Gamma$ (the same parameters as the ones used in panels (b) to (d)
in Fig.~3). For all the cases analyzed, the first crossing of $T_0(V_g)$ and $T_0(V_g+U)$
agrees with $V_{gc}$, and, as can be clearly seen, the second crossing occurs always 
at the same gate potential $V_g/U=-1.0$ [note that in panel (f), the two crossings 
now coincide, resulting in the peculiar charging pattern in Fig.~3(d)].

As initially hinted by the finite temperature NRG results in Fig.~4, the gate potential dependent charge oscillation 
effect is a low-energy many-body effect; in reality, an interesting competition 
between two different many-body states, Kondo and Intermediate Valence. The ability of the DMRG calculated 
$T_0(V_g)$ to explain this subtle effect motivates us to compare $T_0$ with the 
well known Haldane expression for $T_K$, \cite{haldane,david1}
\begin{equation}
k_BT_K \sim D\left(\frac{\Gamma U}{|V_g||V_g+U|}\right)^{1/2}e^{-\pi |V_g||V_g+U|/2\Gamma U}~.
  \label{eq7}
\end{equation}

Fig.~8 does just that for a system consisting of a {\it single} QD. A finite size 
scaling of $T_0$, for DMRG clusters of increasing size, $N=10$, $20$, $30$, and $40$ (where $N$ is the total 
number of sites in a DMRG one-dimensional cluster containing the impurity plus $N-1$ non-interacting sites), 
shows that $T_0$, in the $V_g$ region around the p-h symmetric point, where the Haldane expression
applies, is tending to Haldane's $T_K$. The parameter values are indicated in the figure. 
A few points have to be stressed here. First, note that the objective of Fig.~8 was not to 
attempt a simulation of the absolute value of {\it $T_K$} (all the curves have been translated in energy so 
that they coincide with the $N=40$ value at the p-h point). The objective at this point is to 
reproduce the dependence of $T_K$ with gate potential $V_g$. Second, note that the 
Haldane expression is by no means exact, and is supposed to be applicable only when real 
charge fluctuations of the impurity can be neglected, therefore we are not expecting a complete agreement 
of DMRG results with Haldane's $T_K$. With these two points in mind, and taking into account that 
a larger DMRG computational effort (not attempted in this work) can greatly improve the 
accuracy of the $T_0$ results, one can, after a careful 
analysis of different definitions for $T_0$, propose a DMRG calculation of $T_0$ as a useful guide to experimentalists 
for all the relevant regimes of the single impurity Anderson model. \cite{note4}

\section{Conclusions}

In this work, motivated by recent interest in CNT QD charge transport measurements, the authors 
use three distinct numerical techniques to study 
the 2LSU2 regime and present details of its properties, which hinge in two 
important aspects: first, a transformation from atomic to molecular orbitals 
is needed to understand the numerical results for charge transport. Second, this transformation unveils an 
interesting gate-potential-dependent charge oscillation involving two strongly interacting molecular orbitals, 
one of which, MO $|+ \rangle$, is coupled to the band, while the other, MO $|- \rangle$, 
is totally decoupled from it ({\it at} the 2LSU2 state), while capacitively coupled to $|+ \rangle$. 
In doing that, a series of unanswered questions associated to the 2LSU2 regime are clarified. 
First among them is the fact that {\it at} the 2LSU2 regime the maximum conductance passes from 
$2G_0$ to $G_0$, as the charge transport changes from double- to single-channel. In addition, 
all the charge transport properties specifically associated to the 2LSU2 state are shown 
to be robust, in the sense that they can in principle be observed even when there is no 
total mixing of the orbital quantum numbers ($\phi < \pi/4$). 
Finally, by numerically defining the quantity $T_0$, it is shown that this gate dependent charge oscillation results from 
the competition between the Kondo and Intermediate Valence states. These two states, 
separated by a wide crossover region, can be effectively put in contact by 
the presence of a strongly correlated {\it dark state} (i.e., a zero width state), 
the MO $|- \rangle$. By comparing 
our numerical results for $T_0$ with Haldane's $T_K$, we speculate that the DMRG results obtained by this relatively 
easy computational calculation may be of interest to experimentalists in the analysis of other systems. 
A more refined analysis of $T_0$ is now being pursued and will be presented elsewhere. In addition, 
it will be the subject of a forthcoming work to understand how the phenomena observed in the 2LSU2 regime 
change for $U^{\prime} \neq U$ and when the degeneracy of the AOs $|\alpha \rangle$ and $|\beta \rangle$ is lifted. 

\begin{acknowledgements}
It is a pleasure to acknowledge useful conversations with K.~H. Al-Hassanieh and F.-H. Meisner.  
PO and GAL acknowledge financial support of FONDECYT (Chile) under grants 1080660 and 1100560. 
GBM and CAB acknowledge financial support by the National Science Foundation under Grant No. DMR-0710529, 
and AF acknowledges financial support from NSF Grant No. DMR-0955707. EV acknowledges support 
from CAPES, CNPq, and FAPEMIG (Brazil), and EVA acknowledges financial support ofthe Brazilian agencies 
FAPERJ (CNE) and CNPq (CIAM). 
\end{acknowledgements}

\appendix

\section{Canonical transformations}
\label{AppendixA}

The system studied in this work consists of four different Fermi seas connected to a two-level CNT QD, schematically shown in Fig.~\ref{fig1}(a).
Canonical transformations, which eliminate part of the complexity of the problem,~\cite{BJones,Oguri-trg} 
can be performed over the system's Hamiltonian, given by Eqs.~\ref{eq1} to \ref{eq4}. 
In this Appendix, we briefly discuss these canonical transformations and how to calculate the Green's functions and the conductance.

\subsection{Left-Right symmetry}
The first transformation exploits the left-right symmetry. As a consequence of this symmetry 
(see Eq.~\ref{eq4}), just the symmetric combination of left-right creation and annihilation operators in the $\alpha$ ($\beta$) 
band are coupled to the $\alpha$ ($\beta$) orbital of the CNT QD. This can be easily verified by simply rewriting Eq.\ref{eq4} as
\begin{equation}
H_{\rm hyb} = \sum_{\lambda; \lambda^{\prime} ; \sigma} t_{\lambda \lambda^{\prime}} \left[ d_{\lambda \sigma}^\dagger 
(c_{{\lambda^{\prime}}1\sigma}+c_{{\lambda^{\prime}}-1\sigma}) + H.c.\right],
\label{Aeq1}  
\end{equation}
where $\lambda$ and $\lambda^{\prime}$ can take values $\alpha$ or $\beta$.

Now, constructing the symmetric and anti-symmetric combinations of the $c_{\lambda \pm i\sigma}$ operators (where $\lambda=\alpha$ or 
$\beta$) for all $i \leq 1$, we can define a new basis composed of the following orthonormal states,
\begin{eqnarray}
c_{\alpha i \sigma}^e &=& \frac{1}{\sqrt{2}} \left[ c_{\alpha i\sigma} + c_{\alpha -i\sigma} \right], \label{Aeq2}\\
c_{\alpha i \sigma}^o &=& \frac{1}{\sqrt{2}} \left[ c_{\alpha i\sigma} - c_{\alpha -i\sigma} \right],  \label{Aeq3}\\
c_{\beta i \sigma}^e &=& \frac{1}{\sqrt{2}} \left[ c_{\beta i\sigma} + c_{\beta -i\sigma} \right], \label{Aeq4}\\
c_{\beta i \sigma}^o &=& \frac{1}{\sqrt{2}} \left[ c_{\beta i\sigma} - c_{\beta -i\sigma} \right]. \label{Aeq5}
\end{eqnarray}
Note that this change affects only the leads' fermion operators, keeping the CNT QD operators unchanged.
In this new basis, the symmetric (even) combination will remain coupled to the CNT QD, while the 
anti-symmetric (odd) combination decouples from the system. The Hamiltonian 
$H_{\rm hyb}$ in this new basis takes the simple form,
\begin{equation}
H_{\rm hyb} =  \sum_{\lambda; \lambda^{\prime} ; \sigma} \sqrt{2}~t_{\lambda \lambda^{\prime}} \left[ d_{\lambda \sigma}^\dagger 
c_{{\lambda^{\prime}}1\sigma}^e + H.c.\right].
\label{Aeq6}
\end{equation}
Note that the couplings are now renormalized by a factor $\sqrt{2}$ (the factor $2$ accounts for double-counting).
It is easy to show that the band Hamiltonian $H_{\rm band}$ has its linear chain structure preserved under this transformation,  
as crossing terms between different chains cancel out.
The system in this new basis is depicted in Fig.~\ref{fig1}(b).

\subsection{Using $\alpha$-$\beta$ Symmetry}
As the CNT QD sits at the center of the system (equivalent to site 0) it is not affected 
by the previous transformation. Now we can take advantage of the $\alpha$-$\beta$ symmetry to 
simplify even more the problem. Note that under this new $\alpha$-$\beta$ transformation, the 
many-body terms will be also affected. The proposed transformation consists of using the 
symmetric and anti-symmetric combinations between states $c_{\alpha i \sigma}^e$ and $c_{\beta i \sigma}^e$
[see Fig.~\ref{fig1}(b)], and keeping the decoupled (odd) states untouched. The new 
basis for the even subspace will be (where now $i \geq 1$, i.e., $i$ takes only positive values) 
\begin{eqnarray}
a_{i \sigma} &=& \frac{1}{\sqrt{2}} \left[ c_{\alpha i\sigma}^e - c_{\beta i\sigma}^e \right], \label{Aeq7}\\
b_{i \sigma} &=& \frac{1}{\sqrt{2}} \left[ c_{\alpha i\sigma}^e + c_{\beta i\sigma}^e \right], \label{Aeq8}\\
d_{+ \sigma} &=& \frac{1}{\sqrt{2}} \left[ d_{\alpha \sigma} + d_{\beta \sigma} \right], \label{Aeq9}\\
d_{- \sigma} &=& \frac{1}{\sqrt{2}} \left[ d_{\alpha \sigma} - d_{\beta \sigma} \right], \label{Aeq10}
\end{eqnarray}
where the new operators $d_{+ \sigma}$ and $d_{- \sigma}$ act on the $\alpha$, $\beta$ levels of the CNT QD. 
States associated to operators $d_{+ \sigma}$ and $d_{- \sigma}$ are referred to as molecular orbitals (MO) in the main text.
As in the previous transformation, the band Hamiltonian $H_{\rm band}$ retains its independent double-linear-chain 
structure in this case too. 
It is important to note that the total number of particles in the CNT QD is conserved under this transformation.
For the special case $U^{\prime}=U$, the many-body terms in the Hamiltonian $H_{\rm mb}$ 
just depend on the total number of particles. Therefore, its form will not change under 
this transformation: 
\begin{equation}
H_{\rm mb} = \sum_{\sigma;\lambda=\pm} V_g n_{\lambda \sigma} + \frac{U}{2} \left[ N_T^2 - N_T \right],
\label{Aeq11}
\end{equation}
where $N_T=\sum_\sigma(n_{+\sigma}+n_{-\sigma})$. 

The Hamiltonian describing the connection between the CNT QD and the leads takes the form,
\begin{equation}
H_{\rm hyb} = \sum_\sigma \left[ t_+ ~d_{+\sigma}^{\dagger} a_{1\sigma} ~+~ t_- ~d_{-\sigma}^{\dagger} b_{1\sigma} + H.c.\right],
\label{Aeq12}
\end{equation}
where $t_\pm$ is given by Eqs.~\ref{eq5}~and~\ref{eq6}. 
An schematic of the effect of this transformation is shown in Fig.~\ref{fig1}(c).

It is therefore easy to see that in the fully symmetric case ($t^{\prime \prime}=t^{\prime}$),  
and using that $t_-=\sqrt{2}\left( t^{\prime}-t^{\prime \prime} \right)$, 
the MO $|-\rangle$ will be only capacitively 
coupled to the system, as $t_-=0$. As mentioned in the main text, MO $|-\rangle$ is an electronic 
dark state. This effect is also known in the literature as Dicke effect or 
bound state in the continuous (BIC). In the absence of interactions ($U^{\prime}=U=0$) the  
MO $|-\rangle$ is completely decoupled from the charge reservoirs and appears as a delta 
resonance in the LDOS of the CNT QD, moving with $V_g$. 
For $U^{\prime}=U \neq 0$, this state couples capacitively to MO $| + \rangle$ and contributes to an 
effective gate potential (acting on level $|+ \rangle$) given by
\begin{equation}
V_g^*~=V_g + U~n_-,
\label{Aeq13}
\end{equation} 
as explained in the main section of the manuscript.

For the sake of completeness, the many-body Hamiltoninan for the case $U'\neq U$ is provided in the $|\pm \rangle$ basis
\begin{eqnarray}
H_{\rm mb} &=& H_1 ~+~ H_2 ~+~ H_3 ~+~ H_4\\
H_1 &=& \frac{U'+U}{2} \sum_{\lambda=+,-}  \left(  n_{\lambda \uparrow} n_{\lambda \downarrow} +  
n_{\lambda\uparrow} n_{\bar{\lambda}\downarrow}  \right),
\\
H_2 &=& \frac{U'-U}{2}  \sum_{\lambda=+,-}  \left( c_{\lambda\uparrow}^\dagger c_{\lambda\downarrow}^\dagger 
c_{\bar{\lambda}\uparrow} c_{\bar{\lambda}\downarrow} \right),
\\
H_3 &=& \frac{U'-U}{4} \sum_{\lambda=+,-} \left( S^+_\lambda S^-_{\bar{\lambda}} \right),
\\
H_4 &=& U' \left(~n_{+\uparrow}~n_{-\uparrow} +  ~n_{+\downarrow}~n_{-\downarrow} \right).
\end{eqnarray}
Note that, in the case $U'=U$, $H_2$ and $H_3$ vanish, and we recover Eq.~\ref{Aeq11}. 
In addition, in the general case ($U'\neq U$), MOs $|\pm \rangle$ will not 
only be capacitively connected but also spin-spin correlated, 
with the character of the correlations (ferro or antiferro) being dependent on the relative values of $U$ and $U^{\prime}$. 
This points to a possibly very rich phase diagram, which will be fully studied in a forthcoming 
publication. \cite{note1}

\subsection{Green's functions and conductance}
In the main text, we studied this system using three different numerical techniques that require different 
formalisms to calculate the transport properties. We start by explaining how to calculate the 
conductance using LDECA.

In general, transport properties are related to Green's functions that propagate 
electrons from one lead to another (left to right, for example). For the system described in Fig.~1, the 
conductance can be written as,
\begin{equation}
G = 4 \pi^2 ~|g_0|^2 ~t^4 ~[\rho_0(E_F)]^2 ,
\label{Aeq14}
\end{equation}
where 
\begin{eqnarray}
|g_0|^2 &=& |G_{\alpha-1,\alpha 1}|^2 + |G_{\beta -1,\beta 1}|^2  \nonumber \\ &&
+|G_{\beta -1,\alpha 1}|^2 ~+ |G_{\alpha -1,\beta 1}|^2 ,
\label{Aeq15}
\end{eqnarray}
and $G_{\lambda -1,\lambda^{\prime} 1}$ stands for the Green function evaluated at the Fermi level $E_F$,
\begin{equation}
G_{\lambda -1,\lambda^{\prime} 1} = \langle\langle c_{\lambda^{\prime} 1};c_{\lambda -1}^\dagger \rangle\rangle(\omega=E_F),
\label{Aeq16}
\end{equation}
where we keep the notation used in Eq.~\ref{Aeq2} for the creation/annihilation operators and the 
double bracket $\langle\langle;\rangle\rangle$ is the standard notation for Green's functions.
Note that for simplicity we removed the spin subindex from the later equations. 
As our Hamiltonian does not have spin-flip terms, all terms in the propagators 
in this section have the same spin subindex and summation over repeated 
$\sigma$ indexes is assumed.

Now we want to rewrite Eq.~\ref{Aeq14} using the canonical transformation given by 
Eqs.~\ref{Aeq2}, to~\ref{Aeq5}. By replacing in 
Eq.~\ref{Aeq16} the anti-transformation, we obtain two types of propagators,
\begin{eqnarray}
G_{\lambda -1,\lambda 1} &=& \frac{1}{2}\left\{ \langle\langle c_{\lambda 1}^e;{c_{\lambda 1}^e}^\dagger 
\rangle\rangle - \langle\langle c_{\lambda 1}^o;{c_{\lambda 1}^o}^\dagger \rangle\rangle \right\}, 
\label{Aeq17} \\
G_{\lambda -1,\lambda^{\prime} 1} &=& \frac{1}{2}~ \langle\langle c_{\lambda^{\prime} 1}^e;{c_{\lambda 1}^e}^\dagger \rangle\rangle
\label{Aeq18}.
\end{eqnarray}

LDECA is a numerical technique where Green's functions can be easily evaluated. We solve, 
using LDECA, the system represented in Fig.~\ref{fig1}(b). Then, using 
Eqs.~\ref{Aeq14},~\ref{Aeq15},~\ref{Aeq17}, and \ref{Aeq18}, we can 
calculate the conductance as shown in Fig.~\ref{fig6}.

We use two numerical techniques more, NRG and DMRG, but for the transport properties 
we limit these techniques to analyzing the case $t^{\prime \prime}=t^{\prime}$ ($\phi=\pi/4$). 
For this special case, we use all the symmetries of the system [see Fig.~\ref{fig1}(c)].
Within NRG, we calculate the Green's function propagator at the MO $|+\rangle$. Then, 
using the equations of motion for Green's functions,\cite{Zubarev} we 
left-right anti-transformation, obtaining the necessary propagators to 
evaluate Eqs.~\ref{Aeq14} and~\ref{Aeq15}.
For the DMRG technique we use a different approach. After the canonical transformations, 
using the left-right and $\alpha$-$\beta$ symmetries, we have that just 
MO $|+\rangle$ is connected to a reservoir. Using the fact that the system is a Fermi liquid, 
we can apply the Friedel sum rule, i.e., using the charge at the MO level 
$|+\rangle$, the conductance can be calculated by the equation \cite{enrique-cnd}
\begin{equation}
G = 2G_0~\sin^2\left(\frac{\pi~n_+}{2}\right).
\label{Aeq19}
\end{equation}

This procedure cannot be used if $\phi \neq \pi/4$, the conductance cannot be calculated as just described. However, 
the occupancy of the CNT qD levels $\alpha$ and $\beta$, as well as MO $|+\rangle$ and $|-\rangle$, can still be evaluated using DMRG.

\section{Non interacting case}
\label{AppendixB}

In the absence of interactions ($U=0$), the  physical properties of the system described in Fig.~1, as for instance, the 
local density of states (LDOS) and the conductance, can be calculated exactly.

We can start by analyzing the LDOS at each level of the QD. As we mentioned in Appendix~\ref{AppendixA} a 
canonical transformation can be performed mixing levels $\alpha$ and $\beta$ 
of the QD into two new levels, a symmetric MO $|+\rangle$  and an anti-symmetric 
MO $|-\rangle$. The LDOS at any of the AO levels of the QD can be calculated 
as the superposition of the LDOS at MO $|+\rangle$ and MO $|-\rangle$ (since $\rho_\alpha=\rho_\beta$),
\begin{equation}
\rho_\lambda(\omega) = \frac{\rho_+(\omega)+\rho_-(\omega)}{2},
\label{Beq1}
\end{equation}
with $\lambda=\alpha$ or $\beta$.

\begin{figure}
\centerline{\includegraphics[width=3.65in]{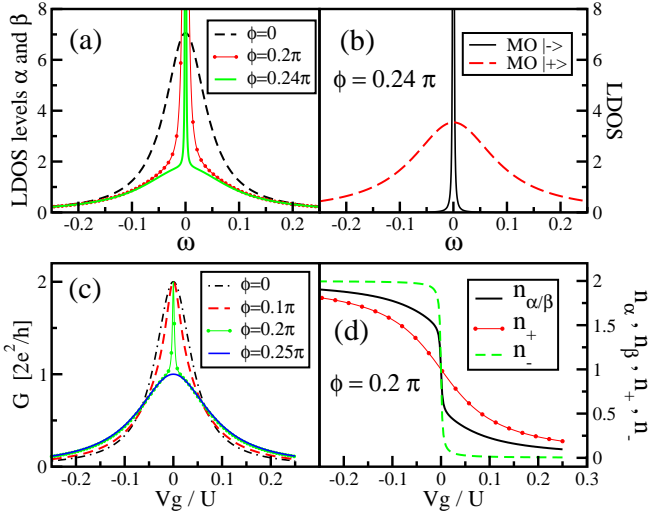}}
\caption{
(a) LDOS of levels $\alpha$ and $\beta$ for $V_g=0$, $t_0=0.15$, and different values of 
$\phi$. When $\phi \to \pi/4$, MO $|-\rangle$ gradually decouples from the band and 
a sharp resonant state develops at $V_g$ in the LDOS. 
(b) LDOS for MOs $|\pm \rangle$ for $\phi=0.24\pi$. 
(c) Conductance as a function of $V_g$ calculated for $t_0=0.15$ and different values of 
$\phi$. When $\phi=\pi/4$ the MO $|-\rangle$  is totally decoupled and the conductance 
reduces by half, indicating that just one channel is active. 
(d) Charging of levels $\alpha$, $\beta$, and MOs $|\pm \rangle$ as a 
function of $V_g$ for the case $\phi=0.2\pi$. As MO $|-\rangle$ is weakly coupled, its charging resembles a step function.
}
\label{Bfig1}
\end{figure}

For $U=U^{\prime}=0$, MOs $|+\rangle$ and $|-\rangle$ are totally independent. 
Their LDOS is associated to the corresponding coupling with the contacts. Each 
of these levels will have a width given by 
\begin{equation}
\Gamma_\delta(\omega)=\pi {t_{\delta}}^2 \rho_0(\omega),
\label{Beq2}
\end{equation}
where $\delta$ stands for $|+\rangle$ or $|-\rangle$ and $\rho_0$ is the LDOS at the leads.
As shown above, in the main text and in Appendix A, when $\phi \to \pi/4$, the anti-symmetric state 
gradually decouples from the charge reservoir (as $t_-\to 0$) and its LDOS becomes a delta function. 
When $\phi=\pi/4$ the MO $|-\rangle$ is completely decoupled and  $\rho_{\alpha}=\rho_{\beta}$ is 
a superposition of a delta function and a resonant peak with width $\Gamma_+$.
A similar effect can be observed in other DQD systems\cite{Guevara1, Pedro2} or in a double Rashba ring\cite{Marcelo1}.

Figure~\ref{Bfig1}(a) shows $\rho_{\alpha}$ and $\rho_{\beta}$ for 
different values of $\phi$. The LDOS was calculated using the Green's function equations of motion formalism, 
which is exact in the limit of $U=0$. 
Fig.~\ref{Bfig1}(b) shows the LDOS for the MOs $|+\rangle$  and $|-\rangle$  
for $\phi=0.24\pi$ and $V_g=0.0$. For this value of $\phi$, close to $\pi/4$, the MO $|-\rangle$ 
is almost decoupled and therefore appears as a narrow resonance located at $V_g$.

The formula for the conductance is given by \cite{enrique-cnd}
\begin{eqnarray}
G(V_g) &=& 2G_0\frac{\tilde{t_0}[t^2V_g^2(1+\sin^2 2\phi) + \tilde{t_0} \cos^42\phi]}
{T_+(V_g)~~T_-(V_g)}, \label{Beq3} \\
&& \nonumber \\
T_\pm(V_g) &=& \tilde{t_0}(1~\pm~\sin2\phi)^2 + t^2V_g^2,
\label{Beq4}
\end{eqnarray}
where $G_0=2e^2/h$ and $\tilde{t_0}=4t_0^4$.

When levels $\alpha$ and $\beta$ are not mixed ($t^{\prime \prime}=0$, i.e., $\phi=0$), we can see that 
the maximum of the conductance occurs for $V_g=0$, taking the value $G=2G_0$, 
indicating that both channels are active.

As seen above, for $\phi=\pi/4$, one of the channels is decoupled and the maximum of the conductance is $G_0$.
For values of $\phi$ close to $\pi/4$ Eq.\ref{Beq4} can be written as a superposition of two Breit-Wigner line shapes,\cite{Guevara1, solis}
\begin{equation}
G(\phi\sim \pi/4) \sim G_0 \left( \frac{\lambda^2_+}{V_g^2+\lambda_+^2} + \frac{\lambda^2_-}{V_g^2+\lambda_-^2}\right),
\label{Beq5}
\end{equation}
where
\begin{eqnarray}
\lambda_+ &=& 4t_0^2/t, \label{Beq6}\\
\lambda_- &=& 2t_0^2/t~\left[1-\sin2\phi\right]. \label{Beq7}
\end{eqnarray}
Using Dicke effect terminology, the first term in Eq.~\ref{Beq5} corresponds to the contribution of the super-radiant state 
(symmetric channel $|+ \rangle$) with a width $\lambda_+$, and the second term corresponds to the 
sub-radiant state (anti-symmetric channel $|- \rangle$) with a width $\lambda_-$.

In Fig.~\ref{Bfig1}(c) we show the conductance for different values of $\phi$. 
For $\phi=0$ both channels are active and have the same width. By increasing $\phi$,  
the MO $|-\rangle$ starts to decouple and this is reflected in a narrower 
line in the conductance superposed on the wider line given by MO $|+\rangle$. 
When $\phi=\pi/4$ the MO $|-\rangle$  is completely decoupled and does not give any 
contribution to the conductance, whose maximum is reduced from $2G_0$ to $G_0$. 

Panel (d) of Fig.~\ref{Bfig1} shows the charging for the case $\phi=0.2\pi$. 
Note that the charging of MO $|-\rangle$  is almost a step function. 
When $\phi=\pi/4$, the charging of levels $\alpha$ and $\beta$ will occur 
discontinuously.



\begin{thebibliography}{99}

\bibitem{saito} R. Saito, G. Dresselhaus, and M.~S. Dresselhaus, 
{\it Physical Properties of Carbon Nanotubes} (Imperial College Press, London 1998).

\bibitem{dekker1} S.~J. Tans, A.~R.~M. Verschueren, and C. Dekker, Nature {\bf 393}, 49 (1998).

\bibitem{lieber} T. Rueckes, K. Kim, E. Joselevich, G. Y. Tseng, C.~L. Cheung, and C.~M. Lieber, Science {\bf 289}, 94 (2000).

\bibitem{hewsonbook} A.~C. Hewson, {\it The Kondo Effect to
   Heavy Fermions} (Cambridge University Press, Cambridge 1993).

\bibitem{kondocnt} J. Nyg\r{a}rd, D.~H. Cobden, and P.~E. Lindelof, 
  Nature {\bf 480}, 342 (2000).

\bibitem{kondosu4} P. Jarillo-Herrero, J. Kong, H. S. J. van der Zant, 
C. Dekker, L. P. Kouwenhoven, and S. DeFranceschi, Nature {\bf 434}, 484 (2005); A. Makarovski, A. Zhukov, J. Liu, and G. Finkelstein,
Phys. Rev. B {\bf 75}, 241407(R) (2007); A. Makarovski, J. Liu, and G. Finkelstein, Phys. Rev. Lett. {\bf 99}, 066801 (2007).

\bibitem{SU4-theo} M.~S. Choi, R. Lopez, and R. Aguado, 
Phys. Rev. Lett. {\bf 95}, 067204 (2005); M.~R. Galpin, D.~E. Logan, and H.~R. Krishnamurthy, J. Phys.:
Condens. Matter {\bf 18}, 6545 (2006); {\bf 18}, 6571 (2006). 

\bibitem{logan} M.~R. Galpin, D.~E. Logan, and H.~R. Krishnamurthy, Phys. Rev.
Lett. {\bf 94}, 186406 (2005).

\bibitem{Aguado} J.~S. Lim, M.-S. Choi, M. Y. Choi, R. Lopez, 
and R. Aguado, Phys. Rev. B {\bf 74}, 205119 (2006). 

\bibitem{martins1} C.~A. B\"usser and G.~B. Martins, Phys. Rev. B {\bf 75}, 045406 (2007). 

\bibitem{eugene} M. Mizuno, E.~H. Kim, and G.~B. Martins, J. Phys.: Condens. Matter {\bf 21}, 292203 (2009).

\bibitem{2lsu2-exp} I.~H. Chan, R.~M. Westervelt, K.~D. Maranowski, and A.~C. Gossard, Appl. Phys. Lett. {\bf 80}, 1818 (2002); 
A.~W. Holleitner, R. H. Blick, and K. Eberl, {\it idem} {\bf 82}, 1887 (2003); 
D.~T. McClure, L. DiCarlo, Y. Zhang, H.-A. Engel, C.~M. Marcus, M.~P. Hanson, and A.~C. Gossard, 
Phys. Rev. Lett. {\bf 98}, 056801 (2007); A. H\"ubel, K. Held, J. Weis, and K.~v. Klitzing, 
{\it idem} {\bf 101}, 186804 (2008); T. Hatanoa, S. Amaha, T. Kubo, S. Teraoka, 
Y. Tokura, J.~A. Gupta, D.~G. Austing, and S. Tarucha, arXiv:1008.0071v1.

\bibitem{first-exp} U. Wilhelm, J. Schmid, J. Weis, and K.~v. Klitzing, Physica E {\bf 14} 385 (2002).

\bibitem{theo} R. Bulla, T.~A. Costi, and T. Pruschke, Rev. Mod. Phys. {\bf 80}, 395 (2008).

\bibitem{white} S.~R. White, Phys. Rev. Lett. {\bf 69}, 2863 (1992).

\bibitem{ldeca} E.~V. Anda, G. Chiappe, C.~A. B\"usser, M.~A. Davidovich, G.~B. Martins, 
F. Heidrich-Meisner, and E. Dagotto, Phys. Rev. B {\bf 78}, 085308 (2008).

\bibitem{martins2} G.~B. Martins and C.~A. B\"usser, Physica B {\bf 403}, 1514 (2008).

\bibitem{martins3} G.~B. Martins, private communication. 

\bibitem{molecular} For an example of this change to molecular orbitals, please see 
E. Vernek, C.~A. B\"usser, G.~B. Martins, E.~V. Anda, N. Sandler, and S.~E. Ulloa, Phys. Rev. B {\bf 80}, 035119 (2009).

\bibitem{haldane} F.~D.~M. Haldane, Phys. Rev. Lett. {\bf 40}, 416 (1978).

\bibitem{note1} Results for a model where inter- and intra-orbital repulsion terms are different will be presented elsewhere.

\bibitem{Pedro1} P.~A. Orellana, M.~L. Ladr\'on de Guevara, F. Claro, Phys. Rev. B {\bf 70}, 233315 (2004).

\bibitem{Pedro2} P.~A. Orellana, G.~A. Lara, and E.~V. Anda, Phys. Rev. B {\bf 74}, 193315 (2006).

\bibitem{Marcelo1} V.~M. Apel , P.~A. Orellana and M. Pacheco, Nanotech. {\bf 19}, 355202 (2008).

\bibitem{Guevara2} M.~L. Ladr\'on de Guevara, P.~A. Orellana, Phys. Rev. B {\bf 73}, 205303 (2006).

\bibitem{solis} B. Sol\'{\i}s, M.~L. Ladr\'on de Guevara, P.~A. Orellana, Phys. Lett. A {\bf 372}, 4736 (2008).

\bibitem{Guevara1} M.~L. Ladr\'on de Guevara, F. Claro, P.~A. Orellana, Phys. Rev. B {\bf 67}, 195335 (2003).

\bibitem{dicke} R.~H. Dicke, Phys. Rev. 89, 472 (1953).

\bibitem{jayaprakash} K. Chen and C. Jayaprakash, Phys. Rev. B {\bf 52}, 14436 (1995).

\bibitem{note6} The NRG calculations for very large values of $U/\Gamma$ (panels (a) and (b) of Fig.~3, 
with $U/\Gamma=400$ and $100$, respectively) are too computationally demanding and the authors did 
not achieve the desired convergence.

\bibitem{note2} Results for different values of $t^{\prime \prime}/t^{\prime}$ can be seen in Fig.~3 of 
Ref.~\onlinecite{martins2}. Note that those results are 
for a different regime than the ones presented here (they are located between the Kondo and Intermediate Valence regimes, 
while some of the present results are deeper into the Kondo regime). Note also that, as $t^{\prime \prime}$ varies, 
$\Gamma$ was not kept constant in those calculations.

\bibitem{meden} C. Karrasch, T. Hecht, A. Weichselbaum, J. von Delft, Y. Oreg, and V. Meden, New Journ. Phys. {\bf 9}, 123 (2007).

\bibitem{note3} An analysis of the variation with $V_g$ of the eigenvalues of $H_{\rm mb}$ 
[eq.~(3)] shows that $V_g/U=-1.0$ marks the degeneracy between the ground states of sectors 
with one and two electrons (CB point), conducive to strong charge fluctuations and 
sequential tunneling.

\bibitem{ferro-PRL} G.~B. Martins, C.~A. B\"usser, K.~A. Al-Hassanieh, A. Moreo, and E. Dagotto, Phys. Rev. Lett. {\bf 94}, 026804 (2005).

\bibitem{ferro-JPCM} See G.-H. Ding, F. Ye, and B. Dong, J. Phys.: Condens. Matter {\bf 21} 455303 (2009), and references therein. 

\bibitem{note5}  Based on that, the authors would feel tempted to loosely refer to $T_0$ as the `many-body energy'.
We refrain from that though, as we feel that the definition of $T_0$ and its calculation have to first be refined and better understood.

\bibitem{david1} D. Goldhaber-Gordon, J. G\"ores, M.~A. Kastner, Hadas Shtrikman, D. Mahalu, and U. 
Meirav, Phys. Rev. Lett. {\bf 81}, 5225 (1998).

\bibitem{note4} Computational efforts are underway to explore the accuracy and physical information contained in this 
simple definition of $T_0$. 

\bibitem{BJones} B.~A. Jones, C.~M. Varma, Phys. Rev. Lett. {\bf 58}, 843 (1987).
K. Ingersent, B.~A. Jones, and J.~W. Wilkins, Phys. Rev. Lett. {\bf 69}, 2594 (1992).

\bibitem{Oguri-trg} A. Oguri, S. Amaha, Y. Nishikawa, T. Numata, M. Shimamoto, A.~C. Hewson, S. Tarucha, 
arXiv:1008.1821.

\bibitem{enrique-cnd} E.~V. Anda, F. Flores, J. Phys.: Condens. Matter {\bf 3}, 9087 (1991).
Y. Meir, N.~S. Wingreen, Phys. Rev. Lett. {\bf 68}, 2512 (1992).

\bibitem{Zubarev} D. N. Zubarev, Usp. Fiz. Nauk. {\bf 71}, 71 (1960) 
[English transl.: Soviet Phys. Usp. {\bf 3}, 320 (1960)].
J.~W. Negele and H. Orland, {\it Quantum Many-Particle Systems} (Perseus Books, Reading, MA, 1998).


\end{thebibliography}
\end{document}